\documentclass{aa}
\usepackage{graphicx}
\usepackage{txfonts}
\begin{document}

\title{Evolution of the stellar-merger red nova V1309 Scorpii: \\
          SED analysis}

\author{R. Tylenda\inst{1} \& T. Kami\'{n}ski\inst{2}}

\institute{Department of Astrophysics, 
       Nicolaus Copernicus Astronomical Center,                   
       Rabia\'{n}ska 8,
       87-100 Toru\'{n}, Poland\\
       \email{tylenda@ncac.torun.pl}
    \and
       European Southern Observatory, Alonso de C\'ordova 3107, Vitacura,
       Santiago, Chile\\
      \email{tkamisk@eso.org}}

\date{Received; accepted}

\abstract
{ One very important object for understanding the nature of 
red novae is V1309 Sco. Its pre-outburst observations showed that, 
before its red-nova eruption in 2008, 
it was a contact binary quickly evolving to the merger of the components. It
thus provided us with a direct evidence that the red novae result from
stellar mergers.}
{We will study the evolution of the post-merger remnant of V1309~Sco
over time.}
{We analyse the spectral energy distribution (SED) of the
object and its evolution with time. From various optical
and infrared surveys and observing programmes carried out with OGLE, HST, VVV, 
Gemini South, WISE, {\it Spitzer}, and {\it Herschel} we constructed
observed SED in 2010 and 2012. Some limited data are also available for
the red-nova progenitor in 2007. We analyse the data with our model of a
dusty envelope surrounding a central star.}
{Dust was present in the pre-outburst state of V1309~Sco.
Its high temperature (900--1000~K) suggests that this was a freshly formed
dust in a presumable mass-loss from the spiralling-in binary. 
Shortly after its 2008 eruption, V1309~Sco became almost completely
embedded in dust. The parameters
(temperature, dimensions) of the dusty envelope in 2010 and 2012
evidence that we then observed matter lost by the object during the 2008
outburst. Its mass is at least $10^{-3}\,M_\sun$. 
The object remains quite luminous, although since its maximum 
brightness in September 2008, it has faded in luminosity by a factor of 
$\sim$50 (in 2012). Far infrared data from {\it Herschel} reveal presence of a
cold ($\sim$30~K) dust at a distance of a few thousand AU from the object.}
{Similarly to other red novae, V1309~Sco formed a slowly-expanding, dense,
and optically-thick dusty envelope during its 2008 outburst. The main
remnant is thus hidden for us. Far infrared
data suggests that the object passed an episode of intense mass loss in its
recent history. This conclusion could be verified by submillimeter
interferometric observations.}


\titlerunning{SED of V1309 Sco}
\authorrunning{Tylenda \& Kami\'nski}
\maketitle
       

\section{Introduction \label{intro}}

The object V1309 Scorpii (V1309 Sco) was discovered as a $\sim$9.5~mag possible nova on
2~September~2008 \citep{nakano}. This relatively faint object did not elicit 
great interest from astrophysicists. It is enough to say that all the available
photometric data for the object obtained during its outburst come from 
amateur astronomers.

Fortunately, \citet{mason10} observed the object spectroscopically. 
Suprisingly enough, these authors
concluded that V1309~Sco was not a classical nova but rather an object 
similar to V838~Mon, which erupted in 2002 \citep{muna02,crause03}. 
Indeed, contrary to classical novae but similarly to V838~Mon, in course of
the eruption, the object evolved to progressively later 
spectral types and declined as a late M-type giant. 
Stellar eruptions showing
this kind of spectral evolution are now usually referred to as red novae or red
transients, although other names, including luminous red novae,
intermediate-luminosity optical transients, intermediate-luminosity red
transients or V838-Mon-type objects, can be found in the literature.

Red novae form a rare class of stellar eruptions. 
Apart from V838~Mon and V1309~Sco, this class in our
Galaxy includes V4332~Sgr \citep{martini} and OGLE-2002-BLG-360 \citep{tku13}. 
There is also growing observational evidence that
CK~Vul (Nova~Vul~1670) \citep{shara85} was a red transient rather
then a classical nova \citep{kato,tku13,kmt15}. Also, V1148~Sgr (Nova~Sgr~1943)
probably belonged to this class, as can be inferred from its spectral
evolution described by \cite{mayall}.

Two red novae have been identified in the Andromeda galaxy, M31\,RV, 
which erupted in 1988 \citep{mould} and the very recent M31LRN\,2015
\citep{williams15,kurtenkov15}. 
A few extragalactic objects, usually referred to as intermediate-luminosity
optical transients, for example M85\,OT2006 \citep{kulk07}, NGC300\,OT2008
\citep{bond09,berger09}, and SN~2008S
\citep{smith09}, could also have been of a similar nature.

Historically, several mechanisms have been proposed to explain 
the red-transient events, including an unusual classical nova 
\citep{it92}, a late He-shell flash \citep{law05}, and a stellar
merger \citep{soktyl03}.
They have been critically discussed in \cite{tylsok06}.
These authors conclude that the merger of two stars is the only mechanism that
can satisfactorily account for the observed properties of red novae. 

Although not best observed in the eruption, V1309 Sco
appeared to be a sort of
Rosetta stone in understanding the nature of red novae. Thanks to the
archive data from the optical gravitational lensing experiment
\citep[OGLE;][]{udal03} it was
possible to follow the photometric evolution of the object over the six years
prior to the outburst \citep{thk11}. The result was amazing: the progenitor
of V1309~Sco was a contact binary quickly losing its orbital angular
momentum and evolving to the merger of the components. Thus V1309~Sco
provided us with a strong evidence that the red-nova events do indeed result
from stellar mergers.

Because of its particular role in the field of red novae, V1309~Sco deserves
careful observational monitoring and thorough study. However,
in just a few years after its eruption the object became
very faint in the optical \citep[see e.g.][]{kmts15}. 
In addition, the object lies in a crowded field.
As discussed in \citet{kmts15} and in the present paper, 
ground-based optical observations now suffer from contamination from
nearby stars. This seriously hampers investigations of the present
state of the object and its evolution. Nevertheless, \citet{kmts15} showed
that V1309~Sco is now in a similar spectral state as V4332~Sgr was a few years after
its 1994 eruption: the optical spectrum of V1309~Sco is now dominated by 
emission features from neutral atoms and molecules. As discussed in 
\citet{kst10} and \citet{tgk15}, this kind of a stellar spectrum can
result from resonant scattering of the central star radiation by atoms 
and molecules in the circumstellar matter, if the central star is
not directly visible. Indeed, in spite of the decline by
about 10 mag in the optical, the object remained bright in the infrared
\citep{nicholls}. Thus the main remnant is now embedded in dust, although not
completely. The atomic and molecular emission features provide evidence that in
certain directions the central star radiation can escape without being
significantly absorbed by dust.

In this paper we investigate the spectral energy distribution (SED)
of V1309~Sco and its evolution with time. Our analysis is based on
broadband photometric data
compiled from various available sources and surveys (Sect.~\ref{obs_sect}). 
We interprete the data with a simple model of a star embedded in a dusty
envelope (Sect.~\ref{model_sect}). From fitting the model results to
the observations we estimate the global parameters of
the object, that is its luminosity, as well as temperature and 
dimensions of the dusty envelope (Sect.~\ref{res_sect}). This is the primary
goal of our study. So far, luminosities of red novae and their
remnants have primarily been derived from optical observations 
\citep[see][for the case of V1309~Sco]{thk11}. 
As we discuss in Sect.~\ref{disc_sect}, this
approach can lead to serious underestimates of the total luminosity and
overestimates of the rate of a red nova decline if the decline is
accompanied with copious dust formation. The latter seems to be a rule
rather than an exception in the red nova population. Hence the need for
observational measurements in a wide range of the infrared wavelengths 
as basic ingredients in analyses of the red nova luminosities. We note that we
are not aiming to investigate detailed dust properties, such as dust grain
composition and dimensions. This would require detailed spectroscopic
measurements in a wide range of the infrared wavelengths as well as a more
sophisticated model of the dusty envelope than that used in our study. An
attempt to analyse the chemical composition of dust grains in V1309~Sco was 
done in \citet{nicholls}.

\section{Observational data  \label{obs_sect}}

\subsection{Herschel observations}\label{hersch_sect}

Observations with the photodetector array camera and spectrometer (PACS)  
on board {\sl Herschel} were executed on 19 March 2012 and arranged with  
two camera setups: ({\it i}) with blue (70\,$\mu$m) and red (160\,$\mu$m),
and ({\it ii}) with green (100\,$\mu$m) and red channels observed 
simultaneously. The observations were made in the mini scan mapping
mode.
For each of the two camera setups, two scans were made with scanning angles
of 70\degr\ and 110\degr. A typical duration of a single scan was 5\,min.
The data were processed and averaged in each band with the standard PACS
pipeline in the Herschel interactive processing environment 
(HIPE, version 10). We used reduction procedures optimised for aperture    
photometry of point sources. Aperture and colour corrections were applied   
to the measured flux densities which are listed in Table\,\ref{hersch_tab}.

\begin{table} \centering
\caption{Fluxes of V1309\,Sco measured with {\sl Herschel} in March
2012.}
\label{hersch_tab}
\begin{tabular}{@{}c@{ }rrc@{ }c c@{ }c@{}}
\hline
Instr.&
\multicolumn{1}{c}{PSF\tablefootmark{a}}&
\multicolumn{1}{c}{Band}&
$\lambda_{\rm eff}$\tablefootmark{b}&
$\Delta\lambda$\tablefootmark{c}&
Flux&
Error\tablefootmark{d}\\
&&&&&density&\\
 & \multicolumn{1}{c}{(\arcsec)} & &  \multicolumn{2}{c}{($\mu$m)}
 & \multicolumn{2}{c}{(mJy/beam)}\\      
\hline\hline
PACS &~~5.5 & blue  ~70\,$\mu$m & ~68.3 & 21.4 &1\,249.9 & 14.7 \\
PACS &~~6.7 & green 100\,$\mu$m & ~97.9 & 31.3 &~~617.3 & 38.5 \\
PACS & 11.5 & red   160\,$\mu$m & 153.9 & 69.8 &~~216.2 & 46.8 \\[7pt]
SPIRE &~18.2 & PSW 250\,$\mu$m & 242.8 & ~67.6 &~~81.3 & 2.5 \\
SPIRE &~24.9 & PMW 350\,$\mu$m & 340.9 & ~95.8 &~~43.4 & 2.9 \\
SPIRE & 36.3 & PLW 500\,$\mu$m & 482.3 & 185.8 &~$<$56.1 &  \\ 
 \hline\end{tabular}
\tablefoot{
\tablefoottext{a}{Full-width at half-maximum of the point spread
function.}
\tablefoottext{b}{Effective wavelength of the band (calculated for 
the Vega spectrum).}
\tablefoottext{c}{Equivalent width of the band.}
\tablefoottext{d}{Approximate photometry error.}
}
\end{table}

{\sl Herschel} also observed V1309\,Sco using the spectral and photometric
imaging receiver (SPIRE) on 1 March 2012. The observations were performed 
simultaneously in three bands, PSW, PMW, and PLW, which are centred at   
about 250, 350, and 500\,$\mu$m, respectively. The small map mode was 
used with five repetitions. Total observing time was of about 15\,min.    
Data were processed in HIPE-10 using the standard SPIRE pipeline for small
maps. The source fluxes were measured using the time-line source extractor
(profile-fitting procedure {\ sourceExtractorTimeline} in HIPE) optimised
for point sources. The flux densities corrected for the pixelisation   
effect and colour-corrected ones are given in Table\,\ref{hersch_tab}. 
The SPIRE maps in all the bands show an extended and diffuse emission
in the field of V1309\,Sco. V1309\,Sco is located in a void (of a radius
between about 0\farcm5 and 2\arcmin) of such emission in the two lower
bands,
 at 250 and 350\,$\mu$m.  At 500\,$\mu$m, however, where the angular 
resolution is lowest, the diffuse emission severely contaminates 
the position of V1309\,Sco and no reliable  measurements of the flux density
were possible except a conservative upper limit
(Table\,\ref{hersch_tab}).

\subsection{Mid-infrared observations \label{mir_sect}}

\begin{figure*}\centering{%
\setlength{\fboxsep}{0pt}%
\setlength{\fboxrule}{1pt}%
\fbox{\includegraphics[width=0.43\textwidth]{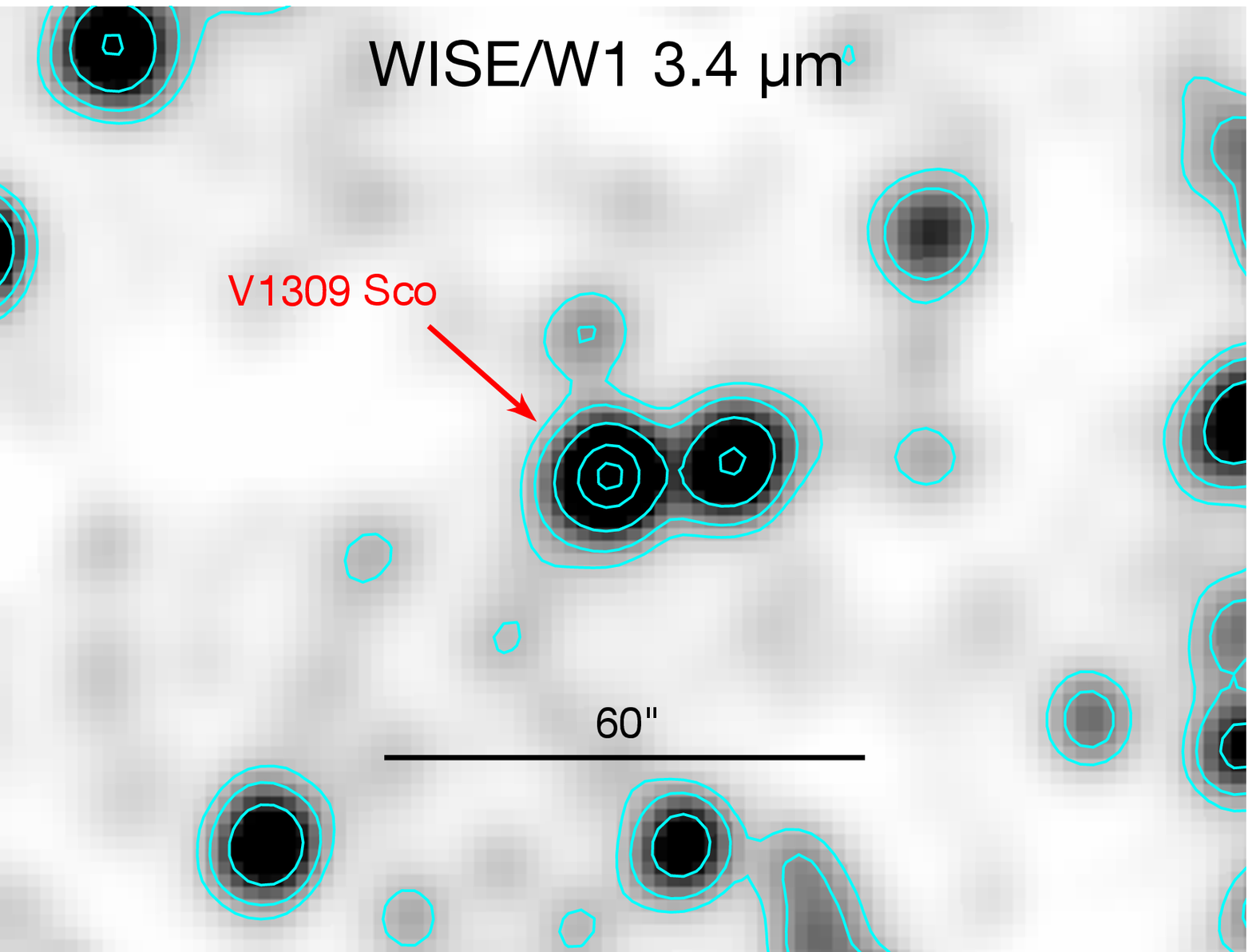}}
\fbox{\includegraphics[width=0.43\textwidth]{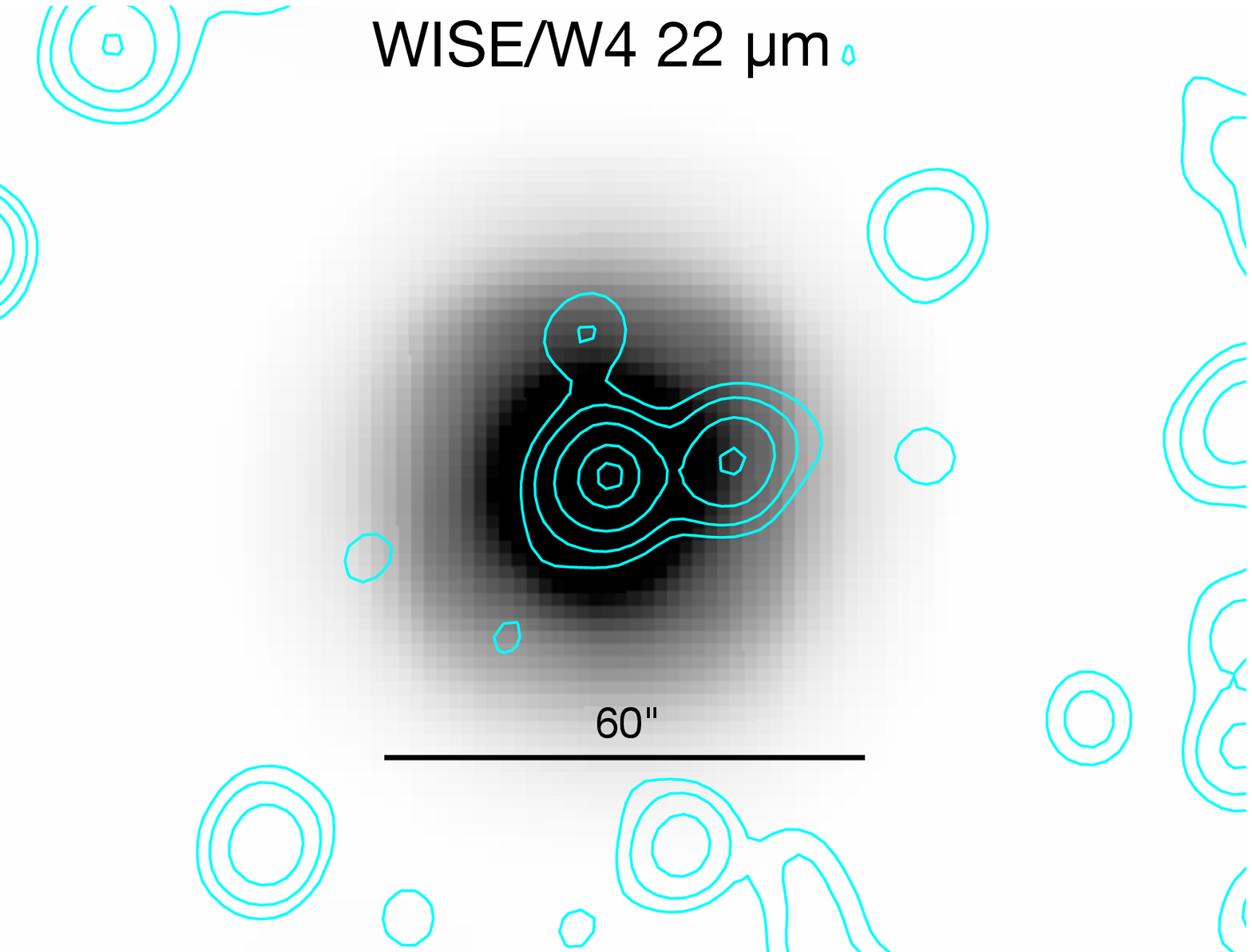}}\\
\fbox{\includegraphics[width=0.43\textwidth]{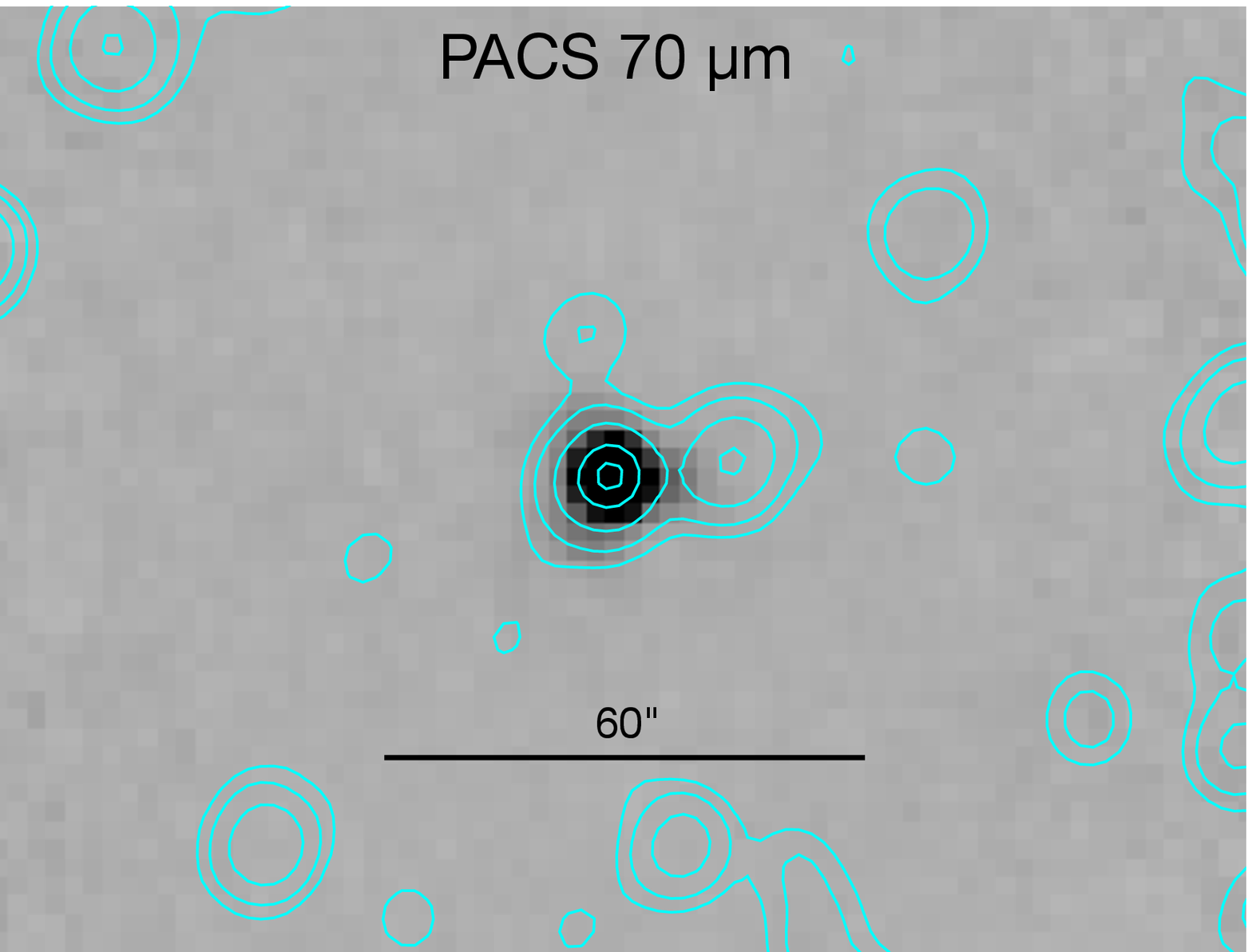}}
\fbox{\includegraphics[width=0.43\textwidth]{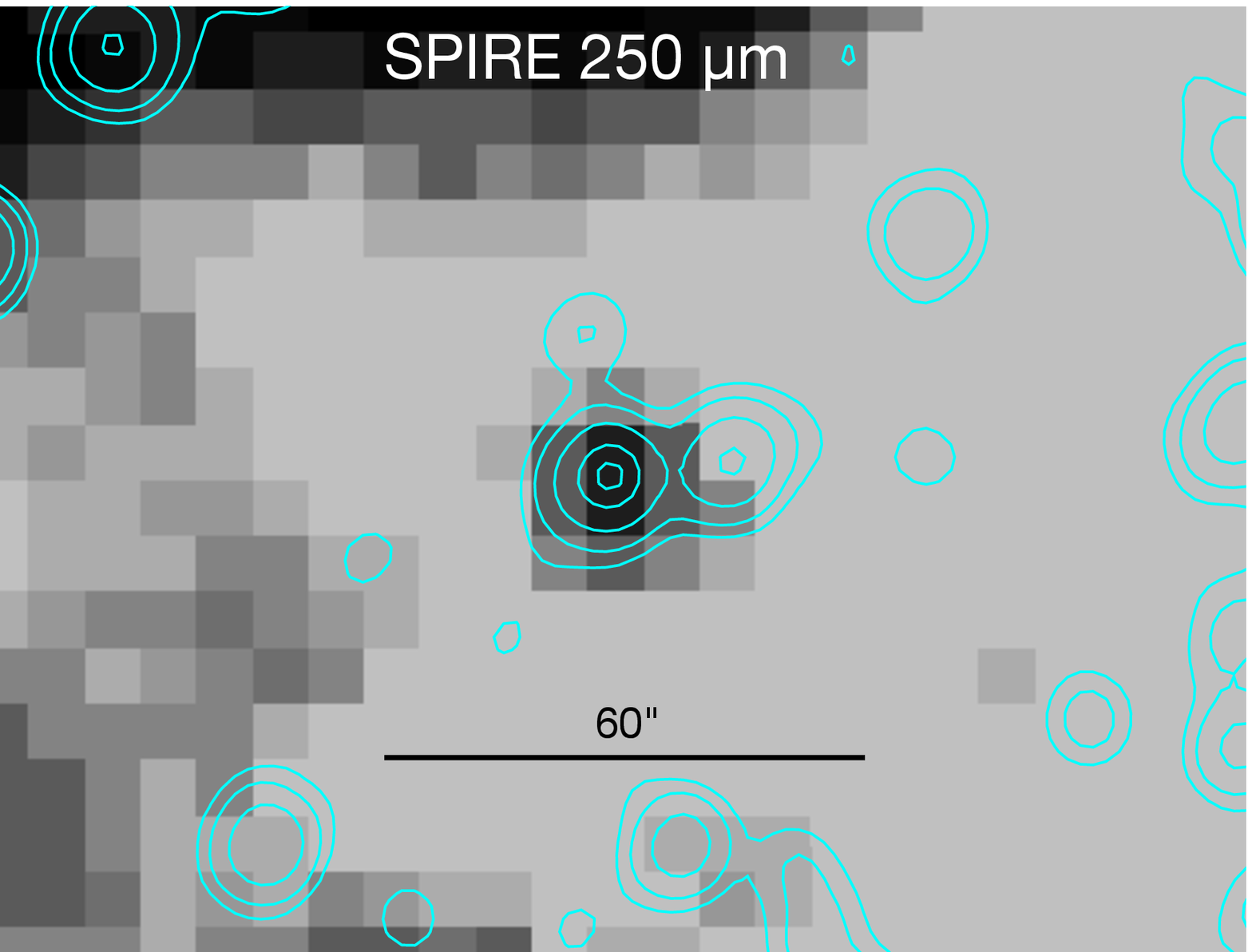}}}
  \caption{Example images of the field around V1309\,Sco in the mid-infrared
and far-infrared. 
The origin and wavelength of each image is given at the top of each panel.
All images are in scale and show the same part of the sky. Contours shown in
all images are drawn for the emission in the W1 band of WISE. East is left, 
north is up.}
\label{image1_fig}
\end{figure*}

The field of V1309\,Sco was covered by several surveys in 
the mid-infrared (MIR) range.
On 19--20 March 2010, the all sky WISE Survey obtained scans
in the four broadbands W1, W2, W3, and W4   
at 3.4, 4.6, 12, and 22\,$\mu$m, respectively. We extracted the survey
images
and source fluxes from the WISE all-sky release Catalog \citep{WISEnew}. 
In the W1 and W2 bands, a blend of at least three sources is seen at the
position
of V1309\,Sco, while at longer wavelengths, the source appears single and    
is dominated by the component that can be identified as V1309\,Sco (see
Fig.~\ref{image1_fig}). 
Our analysis of the WISE images shows that the measurements for 
the source located at the position of V1309\,Sco
are not significantly affected by these nearby sources. 
The WISE magnitudes of V1309\,Sco were converted to monochromatic fluxes
of $W1$=122.88$\pm$2.72\,mJy and $W2$=668.25$\pm$12.93\,mJy, 
$W3$=2.347$\pm$0.024\,Jy, and $W4$=2.883$\pm$0.035\,Jy. These fluxes are
slightly different than WISE fluxes provided in \citet{nicholls}, 
that were probably based on the preliminary version of the Catalog. Our
values agree well with those given in \citet{collum}.

We add to our SED analysis MIR measurements obtained with 
TReCS \citep{trecs-phot} at Gemini south in
August 2010 and as reported in \citet{nicholls}. We found in the Canadian
Astronomy Data Centre photometric observations with the same instrument  
from 6 June 2012 (PI I. Sakon).  
We reduced the data using the standard pipeline in IRAF Gemini package 
and performed aperture photometry. Data were calibrated using at least   
three standard stars in each filter. The results are given in
Table~\ref{ourdat_tab}.

MIR observations at 3.6 and 4.5\,$\mu$m with the {\it Spitzer} Space Telescope
were reported in \citet{collum} for four dates in 2012. We include them in
our analysis. We also add one extra measurement, included in 
Table\,\ref{ourdat_tab}, which was obtained within the same {\it Spitzer} project
on 29 November 2013. To measure the fluxes on all {\it Spitzer} images from 
2012--2013, we used standard pipeline-processed data and performed aperture
photometry (McCollum et al. use profile photometry). The source was 
identified by comparing the field to multi-wavelength images at different
epochs and measured in an aperture of 3\arcsec\ radius. The result was    
corrected for the sky background which was measured as a median within a 
large annulus surrounding the source aperture. An aperture correction was 
applied to the data but no colour correction was employed as it is expected to
be very small. The results are included in Table\,\ref{ourdat_tab}.    
The standard deviation of the measurements at different dates are 2.3\%    
and 15.8\% at 3.6 and 4.5\,$\mu$m, respectively.
Additionally, the data-products we used are expected to have calibration errors
at a level of about 10\%. We note that the MIR fluxes of  
V1309\,Sco, within the errors, did not change considerably over the period
of 2012--2013. We should add that the source we measured covers positions
of several optical stars, including a star $\sim$2\arcsec\ north from V1309\,Sco.      
This star dominates the flux at 3.6\,$\mu$m what is apparent in the
centroid position of the emission region. We were not able to disentangle the 
contributions of the different objects to the total unresolved flux and
therefore the measurements at 3.6\,$\mu$m should be treated as an upper 
level on the actual flux of V1309\,Sco.

\begin{table} 
\centering
\caption{Measured optical, near-IR and mid-IR fluxes of V1309\,Sco.}
\label{ourdat_tab}
\begin{tabular}{cccrrr}
\hline
\normalsize
Origin&
Band&  
$\lambda_c$&
\multicolumn{1}{c}{Date}&
\multicolumn{1}{c}{Flux}&
Error\\  
&&&&\multicolumn{1}{c}{density}&\\
&
&
($\mu$m)&
&
\multicolumn{1}{c}{(mJy)}&
(mJy)\\  
\hline\hline
OGLE & $V$ & 0.55 & 2010 & 0.035 & 0.005 \\
OGLE & $I_c$ & 0.79 & Mar. 2010 & 0.81 & 0.05 \\
OGLE & $I_c$ & 0.79 & Sept. 2010 & 0.53 & 0.05 \\
OGLE & $I_c$ & 0.79 & Mar. 2012 & 0.28 & 0.05 \\
OGLE & $I_c$ & 0.79 & Nov. 2012 & 0.24 & 0.05 \\[7pt]
HST &$F673N$&0.68& 19 Feb. 2012 & 0.04   & 0.01\\
HST &$F160W$&1.54& 19 Feb. 2012 & 2.01   & 0.10\\
HST &$F164N$&1.64& 19 Feb. 2012 & 3.45   & 0.10\\[7pt]
VVV & $Y$   &1.02& 21 May 2010 &3.01 &0.04 \\
VVV & $J$   &1.25& 29 Aug. 2010 &4.48 &0.05 \\
VVV & $H$   &1.65&  9 July 2010 &7.71 &0.07 \\
VVV & $H$   &1.65& 29 Aug. 2010 &6.54 &0.06 \\
VVV & $K_s$ &2.15&  9 July 2010 &8.39 &0.08 \\
VVV & $K_s$ &2.15& 29 Aug. 2010 &6.67 &0.06 \\
VVV & $K_s$ &2.15&  5 Sept. 2010 &6.48 &0.06 \\[7pt]
{\it Spitzer} & ch1 & 3.6 & 28 Oct. 2012 & 5.99\tablefootmark{a} & 10\% \\
{\it Spitzer} & ch1 & 3.6 & 02 Dec. 2012 & 5.94\tablefootmark{a} & 10\% \\
{\it Spitzer} & ch1 & 3.6 & 29 Nov. 2013 & 5.82\tablefootmark{a} & 10\% \\
{\it Spitzer} & ch2 & 4.5 & 28 Oct. 2012 & 62.63 & 20\% \\
{\it Spitzer} & ch2 & 4.5 & 02 Dec. 2012 & 60.66 & 20\% \\
{\it Spitzer} & ch2 & 4.5 & 29 Nov. 2013 & 45.60 & 20\% \\[7pt]
 WISE & W1 & 3.4 & 19 Mar. 2010 &    122.9 & 2.7 \\
 WISE & W2 & 4.6 & 19 Mar. 2010 &    668. & 13. \\ 
 WISE & W3 & 12. & 19 Mar. 2010 & 2\,347. & 24. \\ 
 WISE & W4 & 22. & 19 Mar. 2010 & 2\,882. & 35. \\[7pt] 
Gemini &  Si1 &     7.73  & 6 June 2012 &     313. &    71.  \\
Gemini &  Si3 &     9.69  & 6 June 2012 &     320. &     4.  \\
Gemini &  Si5 &     11.7  & 6 June 2012 &     933. &     5.  \\
Gemini &  Qa  &     18.3  & 6 June 2012 &  2\,487. &    77.  \\
Gemini &  Qb  &     24.5  & 6 June 2012 &  4\,008. &   305.  \\
\hline
\end{tabular}
\tablefoot{
\tablefoottext{a}{The measured flux probably includes some contribution 
from a blending star north from V1309\,Sco.}
}
\end{table}

\subsection{Near-infrared and visual observations  \label{nir_sect}}

\begin{figure*}
\centering{%
\setlength{\fboxsep}{0pt}%
\setlength{\fboxrule}{1pt}%
\fbox{\includegraphics[trim=20 15 25
15,clip=true,width=0.49\textwidth]{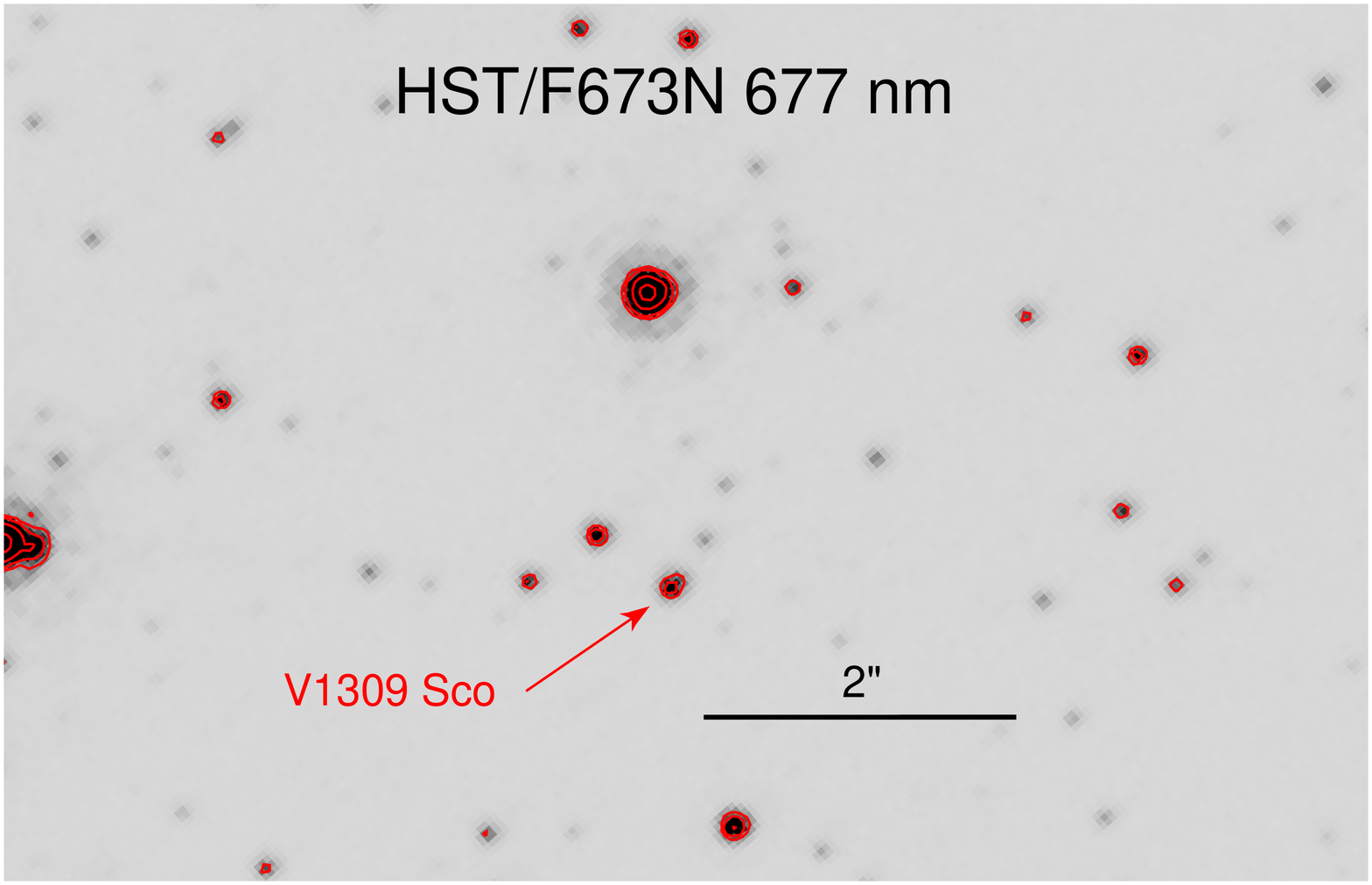}}
\fbox{\includegraphics[width=0.49\textwidth]{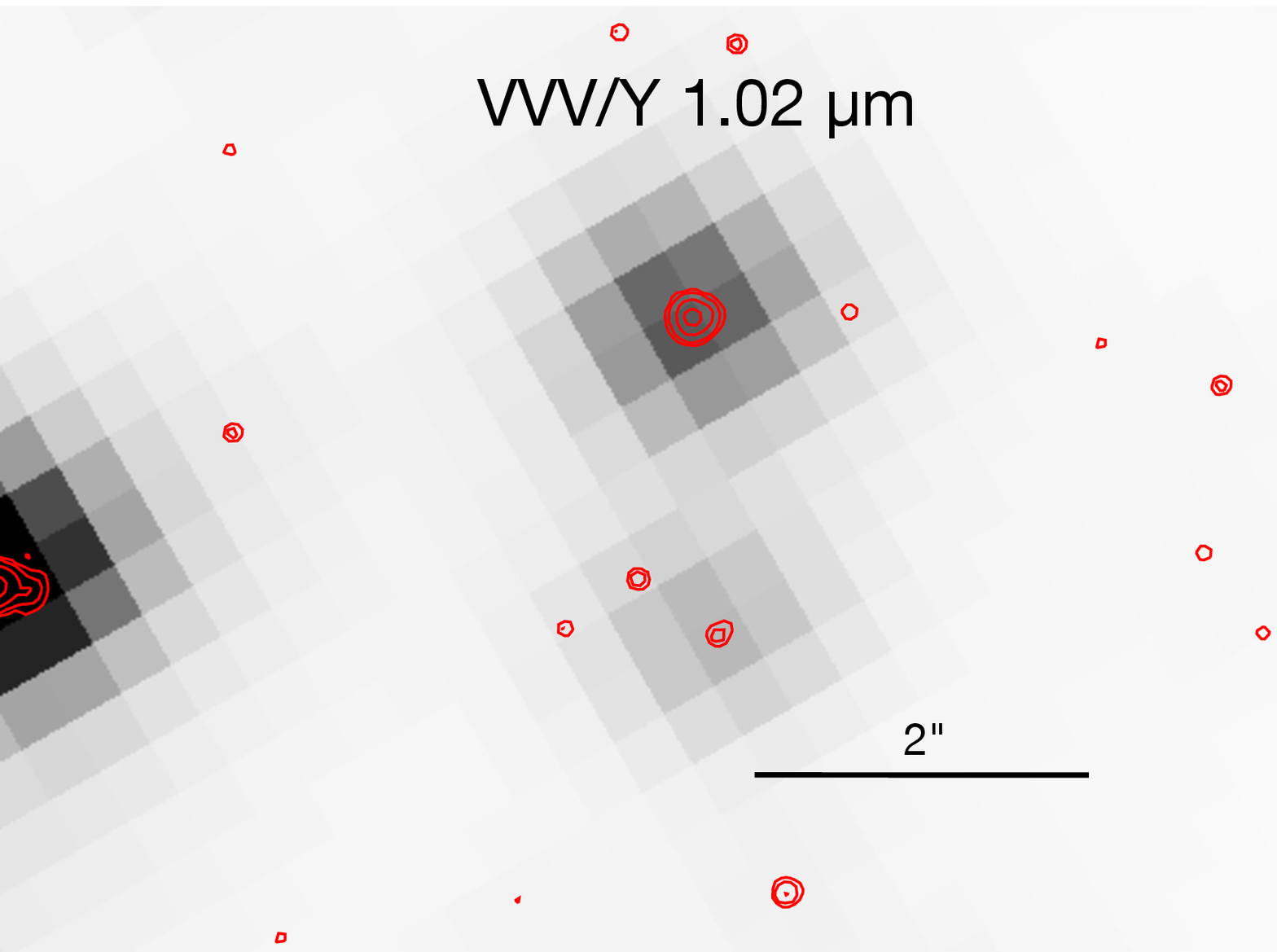}}\\
\fbox{\includegraphics[width=0.49\textwidth]{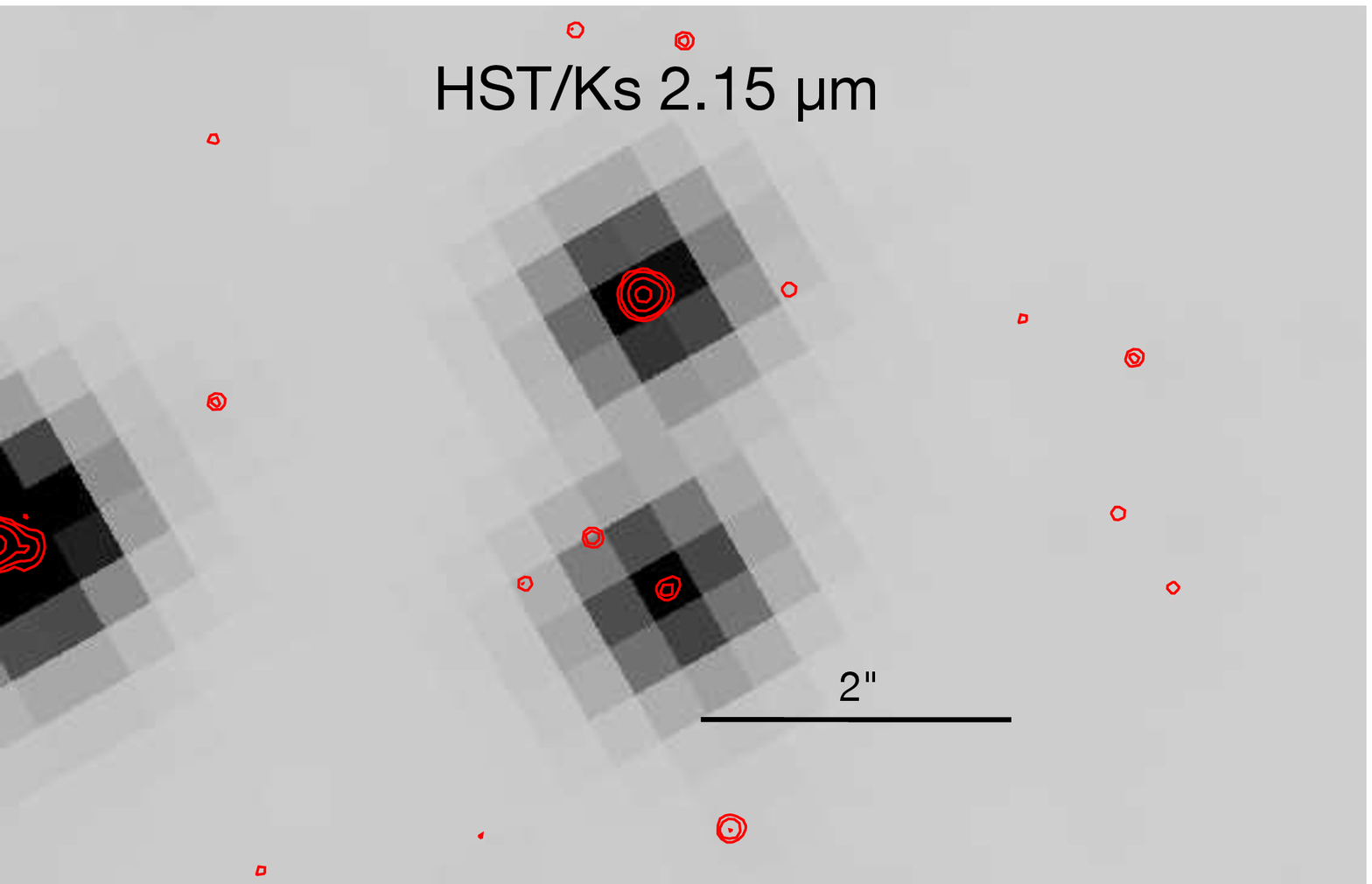}}
\fbox{\includegraphics[width=0.49\textwidth]{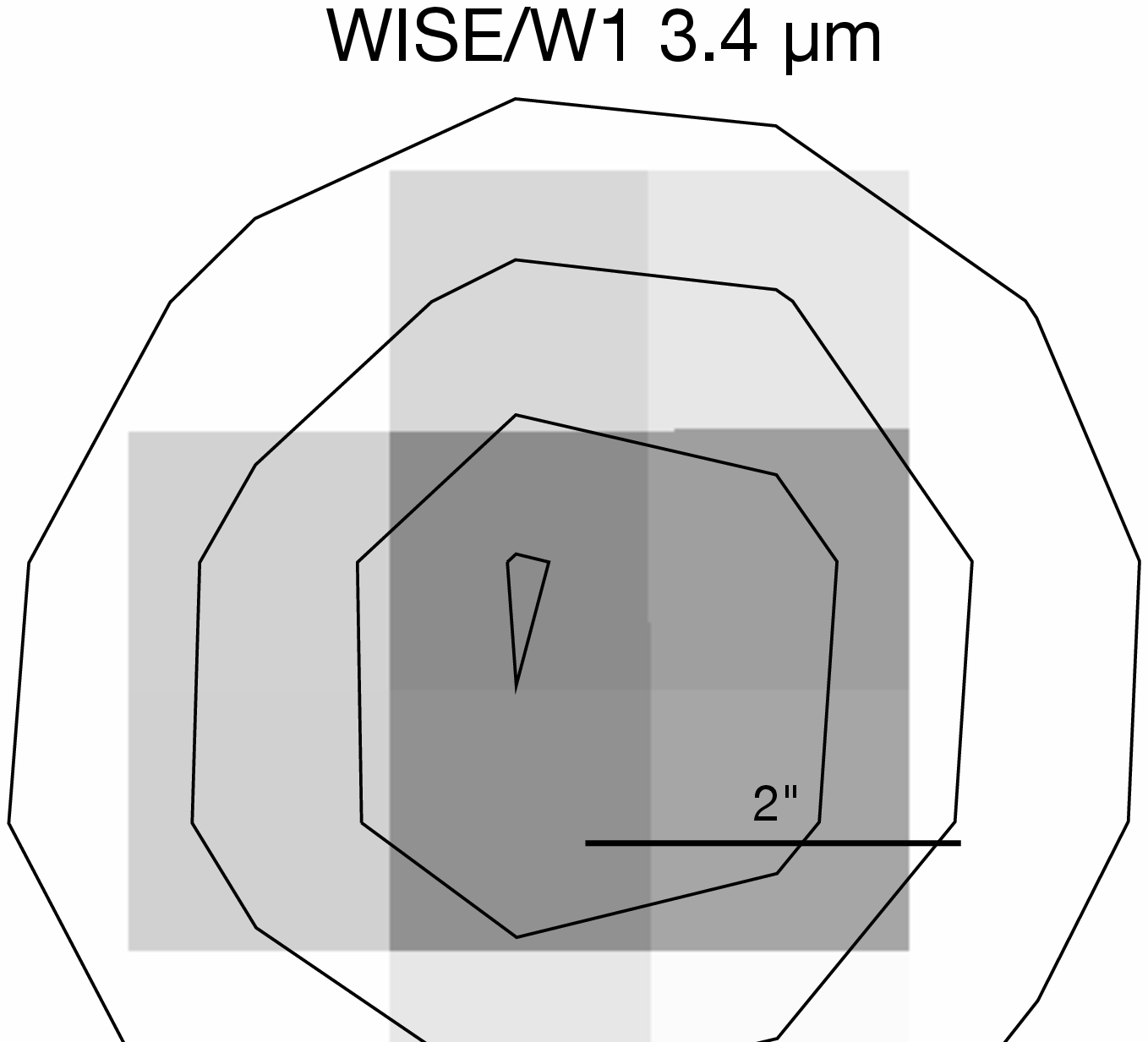}}}
  \caption{Example images of the field around V1309\,Sco at optical and
near-infrared wavelengths. 
All images are to scale and show the same part of the sky. 
The origin and wavelength of each image is given at the top of each panel. 
Contours in all images except that of W1, are shown for the optical emission
in $F673N$. Contours in the WISE image correspond to the emission in 
the W1 band.
 East is left, north is up.}\label{image2_fig}
\end{figure*}

In the near-infrared (NIR) range, the position of V1309\,Sco has been observed within the vista
variables in the Via Lactea (VVV) ESO public survey \citep{vvv} in $ZYJHK_S$
bands. In the first data release of the survey, observations obtained from  
21\,May to 5\,Sept. 2010 are available for this position, at least once in  
each of the five filters. Within the inner part of the PSF of the main source
seen in the W1 band of WISE, two point-like sources are seen. The one 
centred close to the centroid of the emission seen in the W1 band of WISE
is likely to be associated with V1309\,Sco (see below). The other coincident
source is located 1\farcs9 north and partially blends with the component    
associated with V1309\,Sco (see Fig.\,\ref{image2_fig}). VVV catalog entries,  
which include measurements from aperture photometry, in almost all cases,   
provide reliable fluxes for both sources separately, except the $Z$ filter,  
in which the coincident source is considerably brighter than the V1309\,Sco 
component resulting in no measurements for V1309\,Sco in this filter. 
The fluxes for the source associated with V1309\,Sco are given in     
Table\,\ref{ourdat_tab}. 

We also extracted optical and NIR images obtained with the wide field camera 3
(WFC3) of Hubble space telescope (HST) which were obtained on 19-20 February~2012.
They include expositions in the $F160W$, $F164N$, and $F673N$ filters whose 
central wavelengths (widths) are 1.54\,$\mu$m (0.8\,$\mu$m), 1.64\,$\mu$m
(0.7\,$\mu$m), and 0.68\,$\mu$m (42\,\AA), respectively. We analysed the 
standard, pipeline-processed images and performed aperture photometry, which 
was calibrated to flux units ($F_{\nu}$) using the information in the headers 
of the images. Small aperture sizes had to be used in this crowded field and 
therefore all measurements were corrected for the limited aperture size. 
Results of these measurements are included in Table\,\ref{ourdat_tab}. 
The superior angular resolution of the HST images allows us to see that 
the source measured in VVV images, and presumably also in other ground-based
optical data,
is actually a blend of several objects, with two strongest being of comparable
brightness at optical wavelengths. The component that can be
identified as V1309\,Sco is the strongest NIR source among those resolved
sources and marked in Fig.\,\ref{image2_fig}.     

The OGLE monitoring showed a systematic decline of V1309~Sco in 2010--2012
\citep{thk11,kmts15}. Between March and September 2010 the object faded 
from $I \simeq 16.15$ to $I \simeq 16.75$. 
A few OGLE measurements in the $V$ band showed a
roughly constant brightness of $\sim$20.1 mag over the 2010 season. In 2012
the object faded from $I \simeq 17.1$ to $I \simeq 17.5$. No measurement
with the OGLE $V$ filter was available from 2012, but, in this epoch, as
discussed above, 
V1309~Sco was almost certainly dominated by nearby stars in this band.
It is not excluded that the $V$ measurements in 2010 and the $I_c$ values in 
2012 were also contaminated by field stars.
The resultant fluxes from the OGLE observations are given in
Table~\ref{ourdat_tab}.

\section{The model  \label{model_sect}}

The model of a star embedded in a dusty envelope used in the present study is an
updated and modified version of that used in \citet{tku13} to model the SED
of BLG-360. The major modification we have made to create the current version is that the
dust envelope can be composed of up to three concentric shells, and that
the shells are allowed to be truncated at a given elevation angle, 
$\alpha_{\rm trunc}$, above and below 
the equatorial plane. The multiple shells are meant to mimic an extended dust 
envelope with a radial gradient of the dust temperature, as well as to model 
different epochs of mass loss from the object.
The truncation transforms a spherical shell into a torus-like structure.
In this way we can simulate the result of an epoch, when mass loss was
concentrated in the equatorial plane. 

The progenitor of V1309 Sco was an eclipsing binary \citep{thk11}, 
so we assume that the observer is at the orbital plane of the system.

The star, as in \citet{tku13}, is parametrised by its spectral type (or its
effective temperature $T_{\rm star}$) and radius $R_{\rm star}$
(or luminosity $L_{\rm star}$). Its spectrum, $F_{\lambda,{\rm star}}$ is 
interpolated from a set of standard photometric colours,
the same as in \citet{tyl05}. The only difference is that the
spectral type vs. effective temperature calibration was taken from
\citet{levesque} instead of that from \citet{sk82} used in \citet{tyl05}.

\begin{figure}
  \includegraphics[width=\columnwidth]{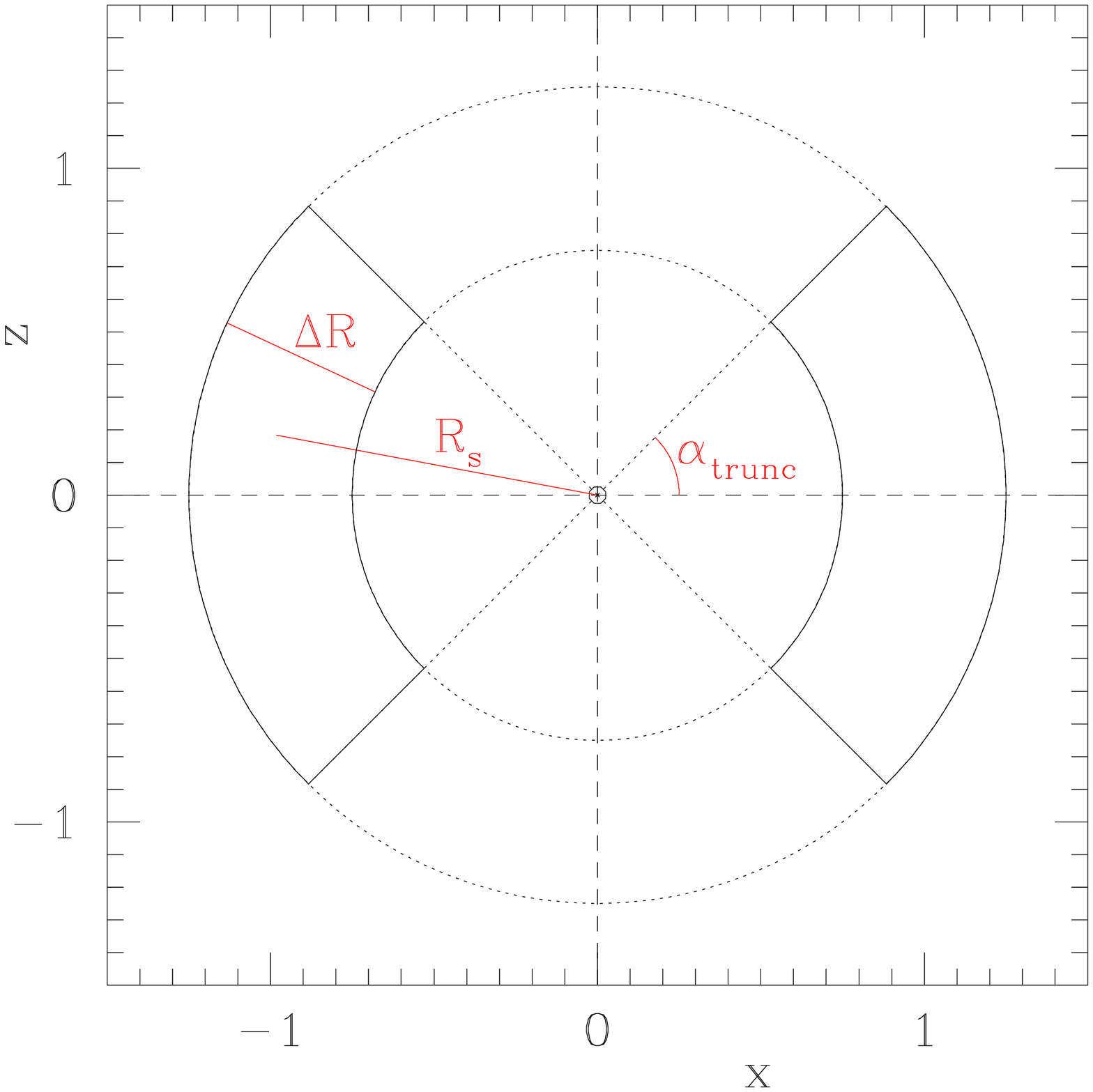}
  \caption{XZ section of the dust shell geometry assumed in our modelling (Y
axis is perpendicular to the figure). 
The shell of a thickness, $\Delta R$, and a radius, $R_{\rm s}$, is illuminated by a source
in its centre. For $\alpha > \alpha_{\rm trunc}$ the shell is truncated, i.e. 
dust is present only within the full lines in the figure. The dust
distribution is symmetrical in regard to the Z axis. The observer is in the XY
plane.}
\label{shell_fig}
\end{figure}

Each dusty shell is spherically symmetric relative to the star and its
geometry is outlined in Fig.~\ref{shell_fig}. The shell is
parametrised by a dust temperature, $T_{\rm s}$,
an optical thickness in the $V$ band along the line of
sight of the central star, $\tau_V^s$, and a relative geometrical thickness of the
shell, $\Delta R/R_{\rm s}$. The latter parameter was taken to be
0.5 in our calculations. The model results are not 
particularly sensitive to this parameter, unless $\alpha_{\rm trunc}$ is small. 
Dust is isothermal and uniformly distributed in the shell.
This last assumption greatly simplifies the radiative transfer in the
envelope - the main reason why we adopted it. It cannot be
justified on physical grounds, but we believe that it introduces smaller
uncertainities in the global parameters of the object, for example total
luminosity, than the unknown geometry of the dusty envelope.

The radius of the dust shell, $R_{\rm s}$, is adjusted to fulfill the
condition of thermal equilibrium, meaning that the shell emits a luminosity
equal to that it absorbs from the incident radiation. In the case of the
innermost shell the incident radiation is just that from the central star.
The outer shell(s) receive(s) a sum of the stellar radiation attenuated by 
dust in the inner shell(s) and the radiation emitted by the inner shell(s),
attenuated if there is a dust shell between the inner emitting shell and the
outer absorbing one. Since the radius of a particular shell is usually much
greater than the stellar radius and the radius(es) of the inner shell(s),
the incident radiation is treated as that from a point source. Even in
the case of models constructed for the progenitor phase (see
Sect.~\ref{prog_sect}), when the exciting source was a binary, this is a
good approximation since the separation of the components
\citep[see][]{thk11} was then at least
an order of magnitude smaller than the radius of the dusty shell.

Spectroscopic observations of V1309 Sco showed numerous molecular bands of
oxides \citep{mason10, kmts15}, which indicates that the object is 
oxygen-rich, that is C/O < 1. Therefore, in our modelling, dust grains are 
assumed to be composed of silicates. Their
optical properties have been taken from the calculations of \citet{dl84}, as
tabulated in the web page of B.\,T. Draine
\footnote{http://www.astro.princeton.edu/{\textasciitilde}draine/dust/dust.html}.
These data refer to the interstellar dust population. However, as thoroughly discussed in \citet{ohm92}, the optical properties of
circumstellar silicates and the interstellar silicates are similar.
Grain sizes follow the standard $a^{-3.5}$ distribution with an upper 
exponential cutoff at 1.0~$\mu$m. The absorption
and scattering coefficients, $Q_{\rm abs}$ and $Q_{\rm sca}$, as well as the
anisotropy factor, $g \equiv \langle \cos \theta \rangle$, are obtained
from integrating the corresponding optical properties over 
the grain-size distribution. 

The incident radiation is absorbed and scattered in a given shell having
an optical thickness, $\tau_\lambda^s$. Thus, the incident spectrum, after
having passed through the shell,
is attenuated by a factor of $\exp (- \tau_\lambda^s)$. 
Outward scattering does not
significantly attenuate the incident radiation, so we define
an effective extinction coefficient as
\begin{equation}
 Q_{\rm ext} = Q_{\rm abs} + Q_{\rm sca} (1 - g).
\end{equation}
 The factor $(1 - g)$ roughly accounts for
the isotropic part of the scattering coefficient.
The optical thickness in front of the star
at a given wavelength, $\tau_\lambda^s$,
is obtained from normalising $Q_{\rm ext}$ to $\tau_V^s$
at the $V$ band. Note that the peak of the silicate feature is at
$\lambda = 9.4\,\mu$m in our dust model and that the optical thickness at 
this wavelength is equal to $0.46\,\tau_V^s$.

In order to calculate the outgoing emission from a given dust shell, we treat
the shell as a circular slab or disc of radius 
$R_{\rm s}+0.5\,\Delta R$,
resulting from projection of the
spherical-shell dust distribution on a two-dimensional plane. At a given
position on the disc, a local monochromatic intensity, $I_\lambda$, 
is then obtained from
\begin{equation}
 \label{i_eq}   
  I_\lambda = B_\lambda(T_{\rm s})\,[1 - \exp\,(-\tau_\lambda)],
\end{equation}
where $B_\lambda$ is the Planck function and $\tau_\lambda$ is the local
optical thickness of the disc. The outgoing monochromatic flux emitted by
the shell is then obtained from integrating $I_\lambda$ over the disc
surface.

If the shell is truncated at a given $\alpha_{\rm trunc}$ above and below 
the equatorial plane, the integration of Eq.~(\ref{i_eq}) is limited to 
the part of 
the disc surface being within the truncation limits. In this case, however,
the shell radiates not only through the outer spherical border but also
through two conical trancution surfaces. Each of them is estimated to have a
surface 
\begin{equation}
 S_{\rm trunc} = 2\,\pi\,R_{\rm s}\,\cos(\alpha_{\rm trunc})\,\Delta R
,\end{equation}
and radiate away a monochromatic intensity
\begin{equation}
\label{ftr_eq}
 I_{\lambda}^{\rm tr} = 
    B_\lambda(T_{\rm s})\,[1 - exp(\tau_\lambda^{\rm tr})],
\end{equation}
where
$\tau_\lambda^{\rm tr} = 
2\,R_{\rm s}\,\sin(\alpha_{\rm trunc})\,\tau_{\lambda}^s/\Delta R$. 
As the observer is supposed to be at the equatorial plane,
the radiation from the truncation surfaces is not included in the
calculations of the observed spectrum.

At shorter wavelengths, scattering dominates
absorption. To account for a random walk of photons in scattering
dominated regions, we defined an effective absorption coefficient as
\begin{eqnarray}
 Q_{\rm abs,eff} & = & \sqrt{Q_{\rm abs} \times Q_{\rm sca}(1- g)}~~~{\rm if}
 ~~~Q_{\rm sca}(1 - g) > Q_{\rm abs}, \nonumber \\
 {\rm or}~~~~~~  & &  \\
 Q_{\rm abs,eff} & = & Q_{\rm abs},~~~{\rm otherwise}.  \nonumber
\end{eqnarray}

The optical thickness of the shell disc in its centre at the effective
wavelength of the $V$ band, $\tau_V^d$, is related to $\tau_V^s$ through 
\begin{equation}
 \tau_V^d = 2\,\tau_V^s \frac{Q_{\rm abs,eff}(V)}{Q_{\rm ext}(V)},
\end{equation}
where $Q...(V)$ stand for respective coefficients taken at the effective
wavelength of the $V$ band. The optical thickness of the disc centre at a
given wavelength, $\tau_\lambda^d$, is then calculated from normalising  
$Q_{\rm abs,eff}$ to $\tau_V^d$ at the effective wavelength of the $V$ band.
The effective thickness at any point of the disc, $\tau_\lambda$,  
scales to $\tau_\lambda^d$ as the dust surface density at the given point to
that at the disc centre.

Finally, we leave a possibility that a certain part of the central star
radiation can be observed as a result of scattering on dust even if the star
is completely blocked in the line of sight. This is a likely situation if
the mass outflow during and after the eruption was not spherically
symmetric. We simulate it with a simple formula,
\begin{equation}
F_{\lambda,{\rm scat}} = f_{\rm scat} \times Q_{\rm sca}(1-g) \times
F_{\lambda,{\rm star}}/4\,\pi
\label{fsc_eq}
,\end{equation}
where $f_{\rm scat}$ is a free parameter in our model.

In summary, the models are specified by the following free parameters:
$T_{\rm star}$, $R_{\rm star}$, $f_{\rm scat}$, and for each dust shell,
$T_{\rm s}$, $\tau_V^{\rm s}$, and $\alpha_{\rm trunc}$.

\section{Results  \label{res_sect}}

The observational data on V1309~Sco described in
Sect.~\ref{obs_sect} can be quite naturally grouped into two epochs, 2010 and 2012.
We supplement 
the data given in Tables~\ref{hersch_tab} and \ref{ourdat_tab} with
measurements published in the literature for these two epochs. When several
measurements are available for a given photometric band, we take a mean
value from them.

To make the
analysis more complete, we also take NIR {\it Spitzer} results obtained in
May~2007 and published in \citet{collum}, and supplement them with optical
measurements from OGLE for the same time period. In this way,
we can also investigate the SED 
in the pre-outburst state of V1309~Sco, although in a rather modest way. 

Following \citet{thk11}, we assume that V1309~Sco is at a distance of 3~kpc
and that it is reddened with $E_{B-V} = 0.8$. 

The observational data from the individual observational epochs are
described in the following subsections and the adopted fluxes are given in
Tables~\ref{prog_tab}, \ref{rem10_tab}, and \ref{rem12_tab}. 
The parameters of the models fitted
to the data are given in Table~\ref{model_tab}. The input parameters are
marked with asterisks. These values were varied to get a good fit to the
observed fluxes. The rest of the parameters in the table are 
the values calculated from the models. 

The upper rows of Table~\ref{model_tab} present the parameters of the
central star, that is the effective temperature and the corresponding spectral
type, the effective radius and the resulting luminosity. All radii
and luminosities given in the table are in the solar units. $f_{\rm scat}$ is a
scattering parameter, which is used in Eq.~(\ref{fsc_eq}) to calculate the
spectrum of the stellar radiation scattered on dust. The next two rows give 
the total observed luminosity of the model and the contribution of the
directly observed radiation from the central star (attenuated by dust along
the line of sight plus the scattered one) to the total observed lumonosity.
The rest of the table presents the parameters of the dust shells, that is the
dust temperature, the optical thickness of the shell in front of the star in
the $V$ band, and the truncation polar angle of the shell (dust is present 
only between $\pm \alpha_{\rm trunc}$ in respect to the equatorial plane). 
The parameters of the shell, calculated in the
models, include the radius, dust surface density (along the line of sight of
the central star), mass of dust in the shell, and the contribution of the
shell to the total observed luminosity.

A comparison of the model spectra with the observed fluxes
is presented in Figs.~\ref{prog_fig}, \ref{rem10_fig}, and \ref{rem12_fig}.
Note that no algorithm for finding the best fit was used. Therefore the
models included in Table~\ref{model_tab} and presented in the figures are
not necessarily best reproducing the observations. The theory used in
getting the models, as described in Sect.~\ref{model_sect}, 
was based on several approximations and simplifications, so we did not want to
exaggerate by using sophisticated fitting procedures. We wanted to get a
certain insight into the evolution of the object, so the figures given in
Table~\ref{model_tab} should be regarded as qualitative estimates of what
was happenning in V1309~Sco during a few recent years.

\begin{table}
\begin{minipage}[t]{\hsize}
\caption{The models (see text for explanations)}
\label{model_tab}
\centering 
\begin{tabular}{ccccc}
\hline
\hline
  Model & 2007\,A & 2007\,B & 2010 & 2012\\[2pt]
\hline
 ($^\ast$) $T_{\rm star}$ & 4700. & 4600. & 3300. & 3000. \\
 Sp.type & K0 & K1 & M6 & M8 \\
 ($^\ast$) $R_{\rm star}/R_\odot$ & 4.7 & 9.5 & 90. & 89. \\
 $L_{\rm star}/L_\odot$ & 9.8 & 36. & 860. & 575. \\[7pt]
 ($^\ast$) $f_{\rm scat}$ & 0.0 & 0.0 & 0.085 & 0.050\\
 $L_{\rm obs}/L_\odot$ & 9.7 &  8.6 & 410. & 220.\\
 $L_{\rm star,obs}/L_{\rm obs}$ & 0.79 & 0.90 & 0.018 
    & 0.012 \\[7pt]
 Shell 1 & & \\
 ($^\ast$) $T_{\rm s}$ & 950. & 900. & 460. & 380. \\
 ($^\ast$) $\tau_V^s$ & 0.35 & 10.0 & 50.0 & 50.0 \\
 ($^\ast$) $\alpha_{\rm trunc}$ & 90\fdg0  & 2\fdg85 & 45\fdg0 & 45\fdg0\\
 $R_{\rm s}/R_\odot$ & 143. & 133. & 3150. & 3770. \\
 $\Sigma_{\rm dust}$[g\,cm$^{-2}$] & $5.0\,10^{-5}$ & $1.4\,10^{-3}$ 
   & $7.1\,10^{-3}$ & $7.1\,10^{-3}$ \\
 $M_{\rm s,dust}/M_\odot$ & $6.3\,10^{-11}$ & $4.9\,10^{-12}$ 
   & $2.5\,10^{-6}$ & $3.6\,10^{-6}$\\
 $L_{\rm s,obs}/L_{\rm obs}$ & 0.21 & 0.096 & 0.66 & 0.28 \\[7pt]
 Shell 2 & & \\
 ($^\ast$) $T_{\rm s}$ & -- & -- & 285. & 180. \\
 ($^\ast$) $\tau_V^s$ &  &  & 4.5 & 15.0\\
 ($^\ast$) $\alpha_{\rm trunc}$ &  &  & 45\fdg0 & 45\fdg0 \\
 $R_{\rm s}/R_\odot$ &  &  & 5040. & 12400. \\
 $\Sigma_{\rm dust}$[g\,cm$^{-2}$] &  &  & $6.4\,10^{-4}$ & $2.1\,10^{-3}$ \\
 $M_{\rm s,dust}/M_\odot$ &  &  & $5.8\,10^{-7}$ & $1.2\,10^{-5}$ \\
 $L_{\rm s,obs}/L_{\rm obs}$ &  &  & 0.33 & 0.69 \\[7pt]
 Shell 3 & & \\
 ($^\ast$) $T_{\rm s}$ & -- & -- & -- & 30. \\
 ($^\ast$) $\tau_V^s$ &  &  &  & 0.23\\
 ($^\ast$) $\alpha_{\rm trunc}$ &  &  &  & 45\fdg0 \\
 $R_{\rm s}/R_\odot$ &  &  &  & 1.2\,10$^6$ \\
 $\Sigma_{\rm dust}$[g\,cm$^{-2}$] &  &  &  & $3.3\,10^{-5}$ \\
 $M_{\rm s,dust}/M_\odot$ &  &  &  & $1.7\,10^{-3}$ \\
 $L_{\rm s,obs}/L_{\rm obs}$ &  &  &  & 0.015 \\
\hline
\end{tabular}
\end{minipage}
(*) Input parameters
\end{table}   

\subsection{The progenitor in 2007  \label{prog_sect}}

V1309~Sco was observed by the {\it Spitzer} Space Telescope IRAC camera on
10~May~2007 in four bandpasses centred at 3.6, 4.5, 5.8, and 7.8 $\mu$m. 
The data were reduced and published by \citet{collum}. 

V1309~Sco was monitored by the OGLE project in the $I_c$
photometric band \citep{thk11}. In May~2007, the mean brightness of the object
was $I = 15.8 \pm 0.2$. The uncertainty comes from the intrinsic
variability of the object at that epoch. There were no OGLE measurements of
the object in the $V$ band during the whole 2007 observing season. We
therefore adopt a $V-I \simeq 2.15$ colour measured by OGLE in 2006
\citep{thk11}.

\begin{figure}
  \includegraphics[width=\columnwidth]{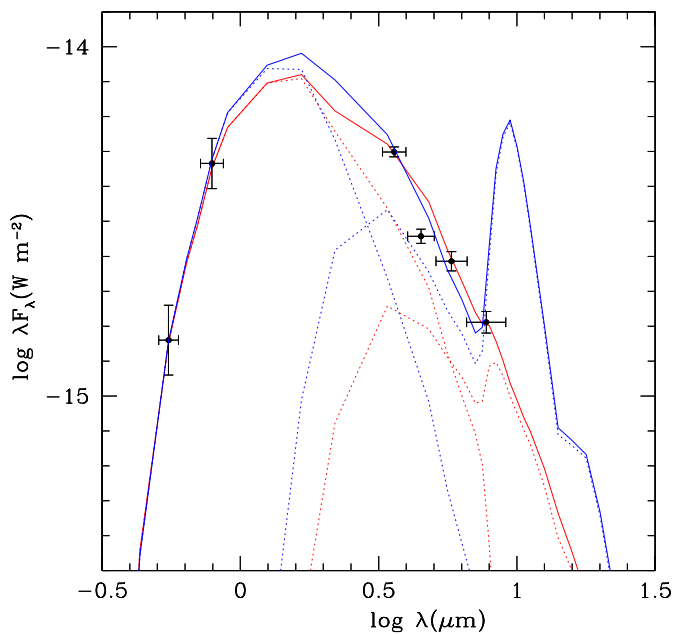}
  \caption{Model fits to the photometric measurements of
the V1309~Sco progenitor in May~2007 (black symbols -- see
Table~\ref{prog_tab}). Blue curves represent  model A. Red curves represent model B.
The parameters of the models are given in Table~\ref{model_tab}.
Dotted curves represent the contributions from the star
(maximum at shorter wavelengths) and the dusty envelope (maximum at longer
wavelengths). Full curves represent the sum of the two components.}
\label{prog_fig}
\end{figure}

The resultant fluxes from the above sources are summarised in
Table~\ref{prog_tab} and plotted as black symbols in Fig.~\ref{prog_fig}.

The observed SED of the V1309~Sco progenitor in 2007 cannot be explained by
a stellar spectrum only. A certain, although not dominating, excess in the
near-IR is clearly present. Fig.~\ref{prog_fig} shows the results of two
models fitted to the observed points. The blue and red curves show model A
and B, respectively, whose parameters are given in Table~\ref{model_tab}.
The dotted curves show contributions from the star (having
maximum at shorter wavelengths) and the dusty envelope (maximum at longer
wavelengths). The full curves display the sum of the two components. Since
we have no observational measurements at wavelengths longer than 10\,$\mu$m,
we have modelled the dust envelope with a single shell. Dust
is relatively hot ($T_{\rm s}=900-950$\,K) in the models. In model A, the shell
is fully spherically symmetric ($\alpha_{\rm trunc} = 90\degr$). The shell
is relatively transparent to the stellar radiation ($\tau_V^s = 0.35$). 
At first sight, the
model spectrum fits well the observational points in Fig.~\ref{prog_fig}.
However, given the width of the {\it Spitzer} 7.8\,$\mu$m bandpass
(2.5\,$\mu$m), the emission in the model 9.4\,$\mu$m silicate band 
seems to be too strong compared to the observed fluxes. Indeed, an
integration of the monochromatic flux from model A over the bandpass of
the {\it Spitzer} 7.8\,$\mu$m filter gives a flux of 7.3~mJy, which is
almost twice as bright as the {\it Spitzer} measurement cited in 
Table~\ref{prog_tab}.

Also the assumption of a uniform, spherically symmetric shell in model A 
does not seem to be reasonable in this particular case. This
assumption requires a spherically symmetric wind from the progenitor.
Assuming an outflow velocity of 100\,km\,s$^{-1}$ (escape velocity from a
1\,M$_\odot$ star with a 5\,R$_\odot$ radius is $\sim$200\,km\,s$^{-1}$), 
the parametres of the model shell would imply a mass-loss rate of the order of 
$10^{-7}{\rm M}_\odot\,{\rm yr}^{-1}$. This is certainly too a high value
for a star of a $\sim$10\,L$_\odot$ luminosity.

As shown in \citet{thk11}, the progenitor was intensively losing
orbital angular momentum, at least since 2002. A part of this process was
very likely due to mass loss via the outer Lagrangian point (L$_2$) 
of the shrinking binary. 
This kind of mass loss is expected to form a geometrically thin disc
at the orbital plane rather than a spherical envelope. Model B (see
Table~\ref{model_tab}) is meant to illustrate this possibility. 
The dust shell is
optically thick in the radial direction ($\tau_V^s = 10.0$) but, in the
vertical direction, the dusty matter fills a disc-like structure with
an opening angle of only $5\fdg7$.
This dusty disc absorbs (and re-emits) only 6.3\% of 
the central star luminosity, but, for an observer in the orbital plane, 
it hides 70\% of the stellar surface. As can be seen from 
Fig.~\ref{prog_fig}, model B reproduces the observations better than model A, 
as it does not produce an intense silicate band in emission.

\begin{table}
\begin{minipage}[t]{\hsize}
\caption{Photometric fluxes for V1309~Sco in May~2007}
\label{prog_tab}
\centering 
\begin{tabular}{llll}
\hline
\hline
 $\lambda$($\mu$m) & $\Delta \lambda(\mu$m) & Flux (mJy) 
 & Source \\[2pt]
\hline
    0.55 & 0.09  &  0.265$\pm$0.06 &  OGLE \\
    0.79 & 0.15  &  1.22~~$\pm$0.2  &  \\[7pt]
    3.6  & 0.7   &  6.0~~~~$\pm$0.2  & Spitzer \\
    4.5  & 1.0   &  4.3~~~~$\pm$0.2  &  \\
    5.8  & 1.5   &  4.7~~~~$\pm$0.3  &  \\
    7.75 & 2.5   &  4.2~~~~$\pm$0.3  &  \\
\hline
\end{tabular}
\end{minipage}
\end{table}   
  
Apart from $\alpha_{\rm trunc} = 90\degr$, which was a priori adopted 
in this model, all the other parameters of model A are rather well 
constrained by fitting the model to the observations. To balance the
infrared excess against the central star spectrum, a decrease of
$\alpha_{\rm trunc}$ requires an increase of $\tau_V^s$. For instance, 
a model with $\alpha_{\rm trunc} \simeq 20\degr$ requires 
$\tau_V^s \simeq 1.0$. To match the observed spectrum, 
the stronger absorption by dust requires that the central star is 
hotter ($\sim$G5 spectral type) and more luminous ($\sim 15\,L_\odot$)
compared to that in model A.

In model B, we adopted $\tau_V^s = 10.0$, in order to suppress the silicate
emission feature at $9.4\,\mu$m. In fact, the feature is significantly
suppressed for $\tau_V^s \ga 5$. In this model, the vertical extension 
of the disc-like shell is smaller than the central star diameter, so the
observed optical flux is dominated by the radiation from the parts of the
star, which are directly observable above and below the disc.
$\alpha_{\rm trunc}$ was adjusted
to balance the infrared dust emission to the observable
stellar radiation. Note that the radiative losses of the shell are here 
dominated by radiation from the truncation sufaces (assumed to be invisible
to the observer). Therefore the results of the model
are somewhat sensitive to the adopted values of $\Delta R/R_{\rm s}$
and $\tau_V^s$. For instance, a model with $\Delta R/R_{\rm s} = 1.0$
and $\tau_V^s = 100.0$ requires $L_{\rm star} = 43\,L_\odot$, 
$T_{\rm s} = 1100\,$K, and $\alpha_{\rm trunc} = 8\fdg3$. The
dust shell is then an order of magnitude more massive but twice as close 
to the star than in model B.

\subsection{The remnant in 2010  \label{rem10_sect}}    

\begin{figure}
  \includegraphics[width=\columnwidth]{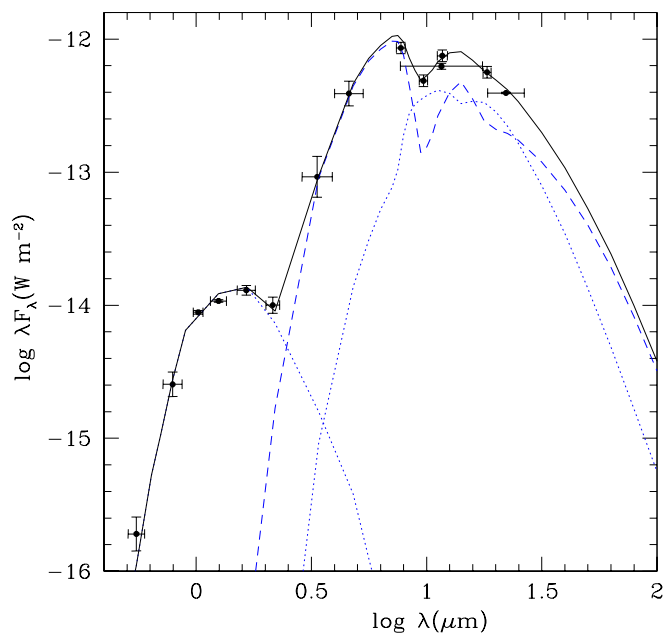}
  \caption{Model spectrum compared to the photometric measurements of
V1309~Sco in 2010 represented with the black symbols 
(see Table~\ref{rem10_tab}).
Dashed and dotted blue curves show contributions of the central star and the
two dust shells to the final model spectrum displayed with the full curve.
Parameters of the model can be found in Table~\ref{model_tab} (fourth column).}
\label{rem10_fig}
\end{figure}

Table~\ref{rem10_tab} presents the fluxes compiled from
various observations of V1309~Sco made between March and September 2010.
Apart from the fluxes given in Table~\ref{ourdat_tab}, we have also taken
into account the results from \citet{nicholls} and \citet{collum}, when
averaging the fluxes for a particular band. The uncertainties of the
resultant fluxes given in Table~\ref{rem10_tab}
include not only estimated precisions of individual measurements but also
discrepancies between the results from different sources.
The fluxes listed in Table~\ref{rem10_tab} are plotted with black symbols in
Fig.~\ref{rem10_fig}.

\begin{table}
\begin{minipage}[t]{\hsize}
\caption{Representative photometric fluxes for V1309~Sco in 2010}
\label{rem10_tab}
\centering 
\begin{tabular}{rccl}
\hline
\hline
 $\lambda$($\mu$m) & $\Delta \lambda(\mu$m) & Flux (mJy) 
 & Source \\[2pt]
\hline
    0.79 & 0.15 &    0.67 $\pm$  0.15 & OGLE, Mar.-Sep. \\
    0.55 & 0.09 &    0.035 $\pm$ 0.005 & \\[7pt]
    1.02 & 0.1~~  &    3.0 $\pm$  0.1 & VVV, May-Sep.\\
    1.25 & 0.2~~  &    4.5 $\pm$  0.1 & \\
    1.65 & 0.3~~  &    7.1 $\pm$  0.6 & \\
    2.15 & 0.3~~  &    7.2 $\pm$  1.0 & \\[7pt]
    7.73 &  0.7~~ &  2210. $\pm$ 220. &  Gemini, Aug. \\
    9.69 &  0.8~~ &  1570. $\pm$  160. & \\
   11.7~~ &  1.2~~ &  2920. $\pm$ 290. & \\
   18.3~~ &  1.5~~ &  3440. $\pm$  340. & \\[7pt]
    3.35 & 1.0~~  &  103.  $\pm$  35. &  WISE, Mar.-Sep. \\
    4.6~~ & 1.3~~  &  600.  $\pm$  120. & \\
   11.6~~ & 9.0~~  &  2420. $\pm$  120. & \\
   22.1~~ & 8.0~~  &  2900. $\pm$  40. & \\
\hline
\end{tabular}
\end{minipage}
\end{table}   
  
As can be seen from Fig.~\ref{rem10_fig}, the SED of V1309~Sco in 2010 was
dominated by infrared emission. The optical and near-IR component, which can
be interpreted as a contribution from the central star, accounts only for
$\sim$2\% of the total observed flux from the object (see
Table~\ref{model_tab}). Clearly the main
remnant of the 2008 eruption is almost completely embedded in dust. The
situation is quite similar to that observed in the remnant of V4332~Sgr 
\citep{kst10}. As shown in \citet{kst10} and
\citet{kt13}, the minor stellar-like contribution observed in V4332~Sgr is
most likely caused by scattering of the central-star radiation on dust
grains in an aspherical, optically thick envelope rather than by a direct 
showing of the central star through a partly translucent envelope. We adopt
that a similar geometry, as in V4332~Sgr, is also observed in the remnant of
V1309~Sco, namely that the dusty envelope is somewhat concentrated near
the equatorial plane, while the central-star radiation can relatively easily
escape along polar directions, giving rise to the scattered
stellar-like component, as well as the atomic and molecular emission features, 
as observed by \citet{kmts15}. Note that the suggested geometry of the dusty
envelope is consistent with what is expected to result from a merger of 
a binary system, that is a dense, relatively slow mass loss concentrated  near
the orbital plane with possible thinner but faster bipolar outflows in polar
directions \citep[similarly to that modelled in][]{akashi}.

The infrared SED observed in 2010 displays a clear 9.4\,$\mu$m silicate
feature in absorption (see Fig.~\ref{rem10_fig}). This cannot be modelled 
by an isothermal dusty envelope. The feature evidently 
shows that the envelope has a temperature gradient with outer cooler dust
superimposed on inner hotter layers forming a sort of an infrared
photosphere. We model this situation with two isothermal shells: the inner
optically thick and hotter shell is surrounded by a cooler one of moderate
optical thickness. Following the above discussion the shells are not
completely spherically symmetric. We adopted that they are 
truncated at $\alpha_{\rm trunc} = 45\degr$. The parameters of our model
can be found in the fourth column of Table~\ref{model_tab}, while the
resultant model spectrum is compared to the observations in
Fig.~\ref{rem10_fig}. The optical thickness of the inner shell
was quite arbitrarily assumed to be $\tau_V^s = 50.0$. In fact the model
spectrum remains practically the same for any $\tau_V^s \ga 20.0$.
The contribution of the model central star to the observed spectrum 
(left blue dotted curve in Fig.~\ref{rem10_fig}) is 
due to scattering on dust grains calculated from Eq.~(\ref{fsc_eq}). The
part of the SED dominated by this contribution, 
 $\lambda \la 2.2\,\mu$m,
was used to estimate the spectral type (effective temperature) of the
central star (see Table~\ref{model_tab}).

As mentioned above, we have adopted $\alpha_{\rm trunc} = 45\degr$ for both
dust shells in our model. This parameter mainly affects the ratio between 
the luminosity of the central star and the observed luminosity of the object
($L_{\rm star}/L_{\rm obs}$). In the presented model this ratio is 2.12 
(see Table~\ref{model_tab}). 
This value partly results from the fact that $\sim$30\%
of the stellar radiation does not interact with the dust shells at all, 
as it escapes through the polar cones limited by 
$\alpha_{\rm trunc}$. The rest is due to radiative losses through the
truncation surfaces of the shells. Obviously, the ratio decreases to 1.0
when $\alpha_{\rm trunc}$ approaches $90\degr$. On the other hand, if we assume 
$\alpha_{\rm trunc} = 25\degr$, $L_{\rm star}/L_{\rm obs}$ increases to 4.8.
From the other model parameters only the radius of the inner dust shell is 
sensitive to $\alpha_{\rm trunc}$, as it varies proprtionally to 
$R_{\rm star}$ ($\sim$$\sqrt{L_{\rm star}}$).

\subsection{The remnant in 2012  \label{rem12_sect}}    

\begin{figure}
  \includegraphics[width=\columnwidth]{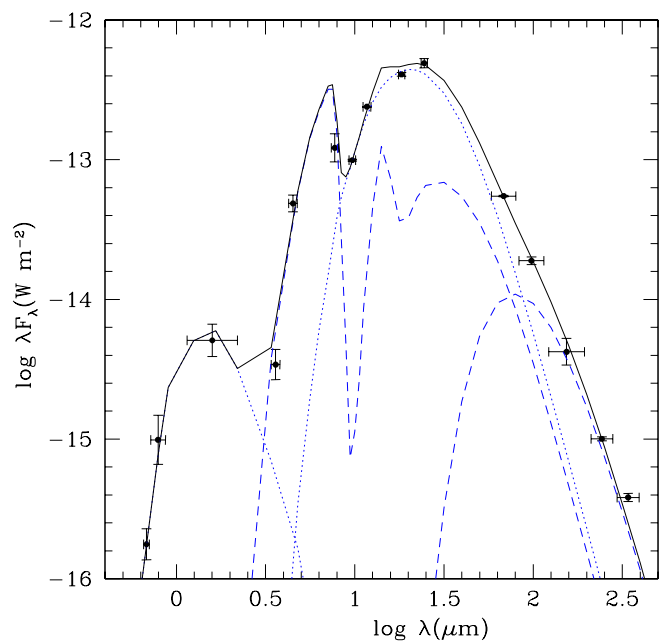}
  \caption{Model spectrum compared to the photometric measurements of
V1309~Sco in 2012 represented with the black symbols 
(see Table~\ref{rem12_tab}). 
Dashed and dotted blue curves show contributions of the central star and the
dust shells to the final model spectrum displayed with the full curve.
Parameters of the model can be found in Table~\ref{model_tab} (fifth column).}
\label{rem12_fig}
\end{figure}

Table~\ref{rem12_tab} presents the fluxes resultant from
various observations of V1309~Sco made in 2012.
The values are primarily based on the fluxes given in Tables~\ref{hersch_tab} 
and \ref{ourdat_tab}. In the case of the data from HST, Table~\ref{rem12_tab}
gives a mean value derived from the
fluxes obtained with the F160W and F164N filters. We have also taken into
account the fluxes obtained by \citet{collum}.
The uncertainties of the
resultant fluxes given in Table~\ref{rem12_tab}
take into account discrepancies between the results for the same band from 
different sources, as well as estimated precisions of individual measurements.
The fluxes listed in Table~\ref{rem12_tab} are plotted with black symbols in
Fig.~\ref{rem12_fig}.

\begin{table}
\begin{minipage}[t]{\hsize}
\caption{Representative photometric fluxes for V1309~Sco in 2012}
\label{rem12_tab}
\centering 
\begin{tabular}{rccl}
\hline
\hline
 $\lambda$($\mu$m) & $\Delta \lambda(\mu$m) & Flux (mJy) 
 & Source \\[2pt]
\hline
    0.79 & 0.15 &    0.26 $\pm$  0.1 & OGLE, Mar.-Nov. \\[7pt]
    0.68 & 0.05 &    0.04 $\pm$  0.01 &  HST, Feb. \\
    1.59 & 1.0  &    2.7 $\pm$  0.7 & \\[7pt]
    3.6~~ &  0.4~~ &    4.1 $\pm$ 1.0 & Spitzer, May-Dec. \\
    4.5~~ &  0.5~~ &   73. $\pm$  10. & \\[7pt]
    7.73 &  0.7~~ &  310. $\pm$ 70. &  Gemini, Jun. \\
    9.69 &  0.8~~ &  320. $\pm$  5. & \\
   11.7~~ &  1.2~~ &  930. $\pm$ 5. & \\
   18.3~~ &  1.5~~ &  2490. $\pm$  80. & \\
   24.5~~ &  2.0~~ &  4010. $\pm$ 300. & \\[7pt]
   68.~~~~ & 21.~~~~ &  1250. $\pm$  15. &  Herschel, Mar. \\
   98.~~~~ & 31.~~~~ &   620. $\pm$  40. & \\
  154.~~~~ & 70.~~~~ &   220. $\pm$  50. & \\
  243.~~~~ & 68.~~~~ &    81.3 $\pm$  2.5 & \\
  341.~~~~ & 96.~~~~ &    43.4 $\pm$  2.9 & \\
\hline
\end{tabular}
\end{minipage}
\end{table}   

As can be seen from Fig.~\ref{rem12_fig}, the SED of V1309~Sco in 2012 is
qualitatively similar to that in 2010 shown in
Fig.~\ref{rem10_fig}. The direct contribution of the central star radiation
in the optical and near-IR remains minor compared to the dominating infrared 
part of the SED produced by dust. The 9.4\,$\mu$m feature is in a deep
absorption.

Our model fits to the observations (see
Table~\ref{model_tab}) however show a
noticeable evolution of the object between 2010 and 2012. The central star
cooled off by $\sim$300\,K and evolved from M6 to an M8 spectral type. The
dusty envelope expanded, as well as, cooled off. Finally, the object faded in
its global luminosity.

The 300\,K drop in $T_{\rm star}$ between 2010 and 2012 
may seem, at first sight, as being too small to be considered as reliablely 
estimated from the observational data. However the
slope of the stellar spectrum in the optical is very sensitive to 
$T_{\rm star}$ in the M-type spectral range. In particular the HST 
($F160\,/\,F673$) colour observed in 2012 is significantly greater 
than that predicted by a model having the same stellar temperature as the
2010 model, that is $T_{\rm star} = 3300$\,K.
Therefore, provided that the scattering properties of dust did not
significantly change in mean time, the derived difference in 
the central star temperature between 2010 and 2012 is realistic.

We have not been able to satisfactorily model the observed SED in 2012 with
two dust shells as we did in 2010. The reason for this is that for 2012 we have far-IR
measurements from {\it Herschel} showing an excess in this region 
(no data from this spectral domain was
available in 2010). A two-shell model fitted to the
optical, near- and mid-IR data in 2012 underestimates the observed far-IR
flux and the difference progressively increases with the wavelength. At
340\,$\mu$m the model underestimates the flux by a factor of $\sim$12.
Therefore, to reproduce the far-IR observations in 2012, 
we have added a third, cold, optically thin and distant shell (see
Table~\ref{model_tab}).
As discussed in Sect.~\ref{disc_dust}, this dust presumably resulted
from a distant-past mass loss from the progenitor, so it must have contributed 
to the observed extinction of the object in 2007 and 2010. As this
contribution was explicitly taken into account in the 2012 model, we have
decreased the interstellar reddening, $E_{B-V}$, from 0.8 to 0.72, when 
fitting the results of this model to the observations.
 
Note that the parameters of Shell~3 are poorly estimated. First of all, 
the shape
of the spectrum in the far-IR suggests that a multi-temperature region would 
fit the data better than an isothermal shell. Second, we adopted the same 
geometry, that is the same value of $\alpha_{\rm trunc}$, for the three shells, 
which need not to be the case. For instance, a model with 
$\alpha_{\rm trunc} = 10\degr$ for
Shell~3 requires $\tau_V^s = 1.0$ and results in $M_{\rm s,dust} = 
3.6\,10^{-4}$\,M$_\odot$. The radius of the shell remains however
practically unchanged.

\section{Discussion  \label{disc_sect}}

The study of the SED of V1309~Sco made in Sect. \ref{res_sect} on the basis
of the observational data described in Sect.~\ref{obs_sect} provided us with
important information on the dusty matter surrounding the object, as well as
on the luminosity of the object. These two subjects are discussed below in
more detail.

\subsection{Dust  \label{disc_dust}}

Dust has dominated the observational appearances of V1309~Sco after its 2008
eruption. It is impossible to understand correctly the evolution of the
object without taking into account dust. 

However, as we show in Sect.~\ref{prog_sect}, dust was also present
in the progenitor phase of the object. 
A possible presence of circumstellar matter in the progenitor of V1309~Sco 
was already considered in \citet{thk11}. 
Those authors suggested that the optical fading by $\sim$1\,mag 
observed in course of a year preceding the 2008 outburst was due
to an increasing mass loss rate from the shrinking binary
progenitor, presumably by its outer Lagrangian point. With the observational
data analysed in Sect.~\ref{prog_sect} we have direct evidence of dust
presence in the progenitor. As argued in Sect.~\ref{prog_sect}, the dusty 
matter was probably concentrated near the orbital plane of the binary
progenitor. The accelerating shrinkage of 
the binary \citep[see Fig.~2 in][]{thk11} most likely resulted in 
an increasing mass loss rate and
a thickening of the dusty disc. In our model 2007\,B, $\sim$70\%
of the photosphere of the progenitor was hidden for us behind the disc. An
increase of the vertical thickness of the disc by 20\% would account for
the observed optical decay of the object by a factor of 2.5
between April~2007 and March~2008 \citep{thk11}.

We can easily see from Table \ref{model_tab} that dust observed in 2010
and 2012 cannot be the same that was seen in the progenitor in 2007. 
First, the dust
mass required to model the SED of the remnant is several orders of magnitude
greater than that in models 2007\,A and B. Second, the estimated radius of
the dusty envelope or disc of the progenitor, $\sim 130\,R_\odot$, 
was significantly smaller
than the effective radius of the object near the maximum brightness in 2008,
$\sim 300\,R_\odot$ \citep{thk11}. Thus dust of the progenitor was
likely to have been destroyed during the eruption. 
Dust seen in the remnant must therefore
have been formed during or after the 2008 eruption. Presumably it took place 
in matter lost during the eruption, when it expanded and sufficently cooled 
off for dust grains to be formed. Following the discussion in
Sect.~\ref{disc_lum}, we can say that dust started
being formed at the beginning of the fast optical decline, that is the end of
September -- beginning of October 2008.

We have used up to three dust shells in our models of V1309~Sco in the
remnant phase in 2010 and 2012. As we discuss below, the first two shells
represents the dusty matter ejected by the object during the 2008 outburst.
The outhermost third shell in the 2012 model probably shows existence of
much older dust in the vicinity of V1309~Sco.

Dividing the dimensions of the two dust shells in our model constructed to
interpret the observations made in 2010 (see Table~\ref{model_tab}) 
by the time elapsed since the 2008 outburst we get velocities of $\sim$40
and $\sim$70\,km\,s$^{-1}$. These values, especially the second one, 
are nicely within the range of
expansion velocities observed during the 2008 eruption \citep{mason10}.
Repeating the same calculations for the inner shells in the 2012 model, 
one gets $\sim$20 and $\sim$70\,km\,s$^{-1}$. 
The apparent deceleration of Shell~1 in the models
can be easily understood in terms of a decreasing opacity in outer
regions of the ejecta due to their expansion. The optically
thick Shell~1 represents dense inner regions where a sort of an infrared 
photosphere is formed. In the expanding envelope 
its position is expected to recede in respect to the outflowing
matter. The partially transparent Shell~2 represents a sort of an
atmosphere in the expanding dusty matter. As the photosphere recedes with
time the mass of the atmospheric regions should increase. Indeed, the mass
of Shell~2 in our models increased by a significant factor between 2010 
and 2012.

The mass ejected by V1309~Sco during its eruption cannot be determined from
our modelling of the SED, as the dusty envelope is evidently optically
thick. The mass of Shell~1 depends on the optical thickness of
the shell. As discussed in Sect.~\ref{rem10_sect}, this parameter is
uncertain since the observed spectrum is insensitive to the optical
thickness if it is greater than $\sim$20.
However a lower limit can be estimated from the dust mass of Shell~2
in the models, as this shell represents observable dust. Adopting the
standard dust-to-gas mass ratio of 0.01, we find from the parameters of the
2012 model (see Table~\ref{model_tab}) that V1309~Sco lost at least 
$10^{-3}\,M_\odot$.

The only measurements of V1309 Sco in the far-IR are those made by 
{\it Herschel} in March~2012 (see Sect.~\ref{hersch_sect}). 
Our model fitted to all the data available in 2012 (see
Table~\ref{model_tab}) suggests existence of a cold dust at a
typical distance of $\sim$5000\,AU. 
It is unlikely that this cold dust could have been produced by the 2008
eruption, because its distance from the central object would imply an
unrealistically high outflow velocity. More probably this matter
resulted from mass loss in the past history of the progenitor of V1309~Sco.
The progenitor was an interacting (contact) binary, whose both
components, according to \citet{stepien}, had already left the main sequence.
One may speculate, following the model of \citet{stepien}, that the moment
when the expanding, more massive component was approaching its Roche lobe,
occured quite recently (in terms of the stellar-evolution time-scale). 
This phase might have beeen accompanied by dynamical
effects related to intense angular momentum transfer between the orbit and
the rotation of the components, which might have triggered an intense mass 
loss from the system.

At a distance of 3\,kpc the dimensions of Shell~3 in our 2012 model would
imply an angular diameter of 3--4$\arcsec$. This should be resolved in
observations using submillimeter interferometers, e.g. ALMA.

\subsection{Luminosity  \label{disc_lum}}

\begin{figure}
  \includegraphics[width=\columnwidth]{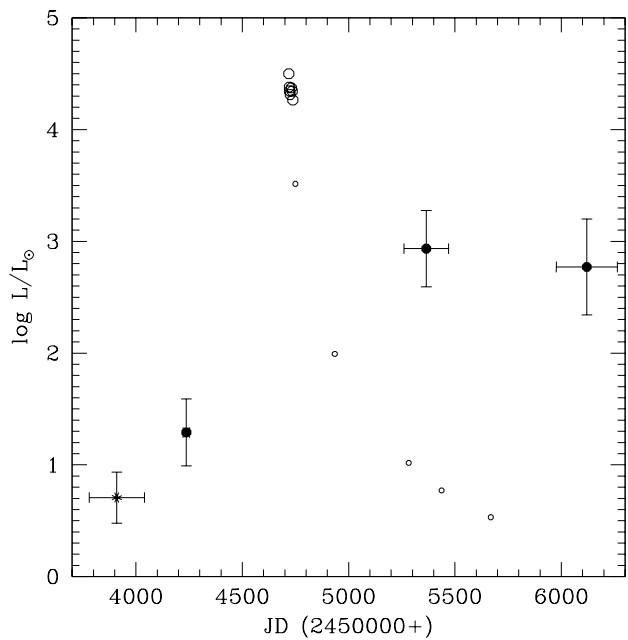}
  \caption{Luminosity of V1309 Sco plotted versus time in Julian Dates. 
Full symbols indicate values derived in the present study. 
The asterisk indicates luminosity of the progenitor estimated in \citet{thk11}. 
Open symbols indicate luminosities during the eruption (big symbols) 
and decline (small symbols)
derived in \citet{thk11}. See text for more explanations.}
\label{evol_fig}
\end{figure}

The estimates of the luminosity of V1309~Sco done in Sect.~\ref{res_sect}
are one of the most valuable results of the present study. 
For a general discussion, we compare them, in Fig.~\ref{evol_fig}, 
with those obtained in \citet{thk11} from optical observations only.

For the progenitor in May~2007 
we obtained two values, depending on which model of the dust
distribution we used. Models A and B, described in
Sect.~\ref{prog_sect} (see Table~\ref{model_tab}), present extreme cases of 
the dust geometry, that is
spherically symmetric shell versus geometrically thin disc. Thus the
resultant luminosites can also be considered as limiting values. 
Therefore the black
symbol at JD\,=\,245\,4240 in Fig.~\ref{evol_fig} shows a mean value from 
these two estimates of 
$L_{\rm star}$ for May~2007 in Table~\ref{model_tab}, while the vertical
error bars show the difference between them.

The results obtained in Sect.~\ref{rem10_sect} and
\ref{rem12_sect} from the observations made in 2010 and 2012 are plotted in
Fig.~\ref{evol_fig} at JD\,=\,245\,5360 and 245\,6120, respectively.
The full symbols in the figure represent $L_{\rm star}$ from
Table~\ref{model_tab}. The values of $L_{\rm obs}$ in the table can be
considered as lower limits to the luminosity of the object. The difference
between $L_{\rm star}$ and $L_{\rm obs}$ is thus taken as an estimate of the
uncertainty of the luminosity of V1309~Sco.

The leftmost asterisk symbol in Fig.~\ref{evol_fig} refers to an estimate
of the luminosity of the V1309~Sco progenitor in 2006 made in \citet{thk11}.
This estimate is model dependent, as it was done assuming that 
the binary components of the progenitor were filling their Roche lobes. In
fact this luminosity estimate
was used in \citet{thk11} to estimate the distance to the object 
($\sim$3\,kpc, a value also adopted in the present study). 
When comparing to our
analysis of the May~2007 data, one can conclude that the V1309~Sco progenitor
brightened by a factor of approximately four between 2006 and 2007. The optical light
curve presented in \citet{thk11} shows that the object indeed brightenned
between these dates but by $\sim$0.35\,mag, that is a factor of $\sim$1.4, only.
The reason for this discrepancy is that we attempted to account for the
observed infrared excess in May~2007. Unfortunately, the result is sensitive 
to the adopted model of dust distribution.
A fit of a pure stellar spectrum (no
dust contribution) to the $V$ and $I$ fluxes in Table~\ref{prog_tab} gives
$L_{\rm star} = 7.1\,L_\odot$, which is 1.42 times the luminosity derived in
\citet{thk11}. As discussed in Sect.~\ref{disc_dust}, the role of dust in
the progenitor increased with time, affecting more and more
importantly the optical appearances of the object, when it was approaching
the merger. Therefore the rise of the object in the bolometric luminosity
during the final years before the 2008 eruption was probably steeper than that of 
the optical brightness observed in \citet{thk11}. As discussed in
\citet{thk11}, dissipation of the orbital energy of the shrinking binary can
easily account for the brightening of the V1309~Sco progenitor, even if
this brightening was more significant than that directly inferred from the
optical lightcurve.

Figure~\ref{evol_fig} shows the luminosities derived in
\citet{thk11} from optical measurements. The phase
of maximum brightness of the object observed between 9 and 28~September 2008, and the decline after the 2008 eruption are both indicated. As is clearly
seen from Fig.~\ref{evol_fig}, the optical data seriously overestimate the
decline rate of the object. The optical decline was largely due to formation
of dust. When the infrared data are taken into account we
have to conclude that a few years after the eruption V1309~Sco is still
relatively bright. It has significantly faded in meantime but between two and three orders of
magnitude less than inferred from the optical data in \citet{thk11}.
From the low fading rate observed between 2010 and 2012 we
can anticipate that the object will remain luminous for many decades. 

The luminosity evolution of V1309~Sco is in fact similar to what is observed 
in other red novae. V4332~Sgr, very similarily to V1309~Sco, developed a huge
infrared excess after its 1994 eruption \citep{kst10}. When this is taken
into account, one can conclude that since the eruption V4332~Sgr has faded 
by a factor of $\sim$30 rather than 1500 as was inferred from the optical data 
only in \citet{tcgs05}. 
The infrared excess of V838~Mon is much less dominant
than those of V4332~Sgr and V1309~Sco, so the optical data give more reliable
estimates of the luminosity. A recent study of \citet{tks11} shows
that in 2009 V838~Mon was a factor of $\sim$60 fainter than at its maximum 
in 2002. Our present study shows that in 2012 V1309~Sco was some 
50 times less luminous than at its maximum in 2008.

\section{Conclusions}

Shortly after its 2008 eruption, V1309~Sco formed a slowly-expanding, dense,
and optically-thick dusty envelope. Its mass is at least $10^{-3}\,M_\sun$.
As a result of dust formation, the optical decline was much 
faster and deeper than that of the bolometric luminosity of the object.
The main remnant is now hidden for us. The cases of V4332~Sgr and CK~Vul 
tell us that it may remain invisible for decades or even hundreds of years. 
In 2012, V1309~Sco remained quite luminous, although since its maximum 
brightness in September 2008, it had faded by a factor of $\sim$50.

Dust was also present in the pre-outburst state of the object.
Its high temperature (900--1000~K) suggests that this was a freshly formed
dust. This was indicative of mass loss from the spiralling-in 
binary, presumably by its outer Lagrandian point, $L_2$. 

 Far infrared data from {\it Herschel} reveal the presence of a
cold ($\sim$30~K) dust at a distance of a few thousand AU from the object.
This suggests that the object passed an episode of intense mass loss in its
recent history. Submillimeter interferometric observations would be able to
verify this conclusion.

\begin{acknowledgements} 

We are greatful to the referee (A. Evans), whose comments helped us
to significantly improve the quality of the paper.

This publication makes use of data several products: 
from the NASA/IPAC Infrared Science Archive, which is operated by the Jet
Propulsion Laboratory, California Institute of Technology, under contract
with the National Aeronautics and Space Administration (NASA); 
from the Wide-field Infrared Survey Explorer, which is a joint project of
the University of California, Los Angeles, and the Jet Propulsion 
Laboratory/California Institute of Technology, funded by NASA;
from the Cambridge Astronomical Survey Unit;
from the NASA/ESA Hubble Space Telescope, obtained from the data archive at
the Space Telescope Science Institute. STScI is operated by the Association
of Universities for Research in Astronomy, Inc. under NASA contract NAS
5-26555;
from the Gemini Observatory (projects GS-2010B-C-7, GS-2011B-C-4, GS-2012A-C-5, acquired through the Gemini Science Archive and processed using the Gemini IRAF package), which is operated by the Association of Universities for Research in Astronomy, Inc., under a cooperative agreement 
with the NSF on behalf of the Gemini partnership: the National Science Foundation 
(United States), the National Research Council (Canada), CONICYT (Chile), the Australian 
Research Council (Australia), Minist\'{e}rio da Ci\^{e}ncia, Tecnologia e Inova\c{c}\~{a}o 
(Brazil) and Ministerio de Ciencia, Tecnolog\'{i}a e Innovaci\'{o}n Productiva (Argentina).
This research used the facilities of the Canadian Astronomy Data Centre
operated by the National Research Council of Canada with the support of 
the Canadian Space Agency.

We acknowledge all these institutions.

\end{acknowledgements}

\end{document}